\documentclass[10pt,letterpaper]{article}
\usepackage[top=0.85in,left=2.75in,footskip=0.75in,marginparwidth=2in]{geometry}

\usepackage[utf8]{inputenc}
\usepackage[T1]{fontenc}

\usepackage{cite}

\usepackage{nameref,hyperref}

\usepackage{microtype}
\DisableLigatures[f]{encoding = *, family = * }

\raggedright
\usepackage{changepage}

\usepackage[aboveskip=1pt,labelfont=bf,labelsep=period,singlelinecheck=off]{caption}
\usepackage{subcaption}
\usepackage{float}

\makeatletter
\renewcommand{\@biblabel}[1]{\quad#1.}
\makeatother

\usepackage{lastpage,fancyhdr,graphicx}
\usepackage{epstopdf}
\fancyheadoffset[L]{2.25in}
\fancyfootoffset[L]{2.25in}

\usepackage{xcolor}
\definecolor{Gray}{gray}{.25}

\usepackage{sidecap}
\usepackage{wrapfig}
\usepackage[pscoord]{eso-pic}
\usepackage[fulladjust]{marginnote}
\reversemarginpar

\usepackage{amsmath,amssymb,amsfonts}

\usepackage{textcomp}
\usepackage{float}
\usepackage{url}

\begin{document}
\vspace*{0.35in}

% Title
\begin{flushleft}
{\Large \textbf{Low-Cost Home Automation System for Municipal Swimming
Pool: Arduino-Based Implementation and Data Analysis}}
\newline
\\
J\'{u}lio Rocha\textsuperscript{1},
Salviano Soares\textsuperscript{1},
Carlos Quental\textsuperscript{2}
\\
\bigskip
\textbf{1} UTAD -- University of Tr\'{a}s-os-Montes and Alto Douro,
Vila Real, Portugal.
\texttt{\{al83013, salblues\}@utad.pt}
\\
\textbf{2} CISeD -- Research Centre in Digital Services, Polytechnic
University of Viseu, Portugal.
\texttt{\{quental\}@estgv.ipv.pt}
\end{flushleft}

% ---------------------------------------------------------------
\section*{Abstract}

This paper presents a low-cost home automation system implemented in a
municipal swimming pool to address various challenges, including security
concerns, air quality control, gas leakage detection, energy consumption
reduction, and temperature and humidity control on the pool deck. The
system utilises Arduino microcontrollers with sensors and actuators,
enabling real-time data collection and analysis. The project is divided
into two phases: hardware assembly and data analysis. In the hardware
assembly phase, the Arduino sends data to a web Application Programming
Interface (API) and stores it in a time-series database, with results
presented in an Android application. The data analysis phase involves
statistical exploration using libraries such as Pandas, NumPy, and
Matplotlib. The proposed system aims to enhance decision-making based
on collected and analysed data.

\begin{figure}[ht]
\centering
\includegraphics[width=0.8\textwidth]{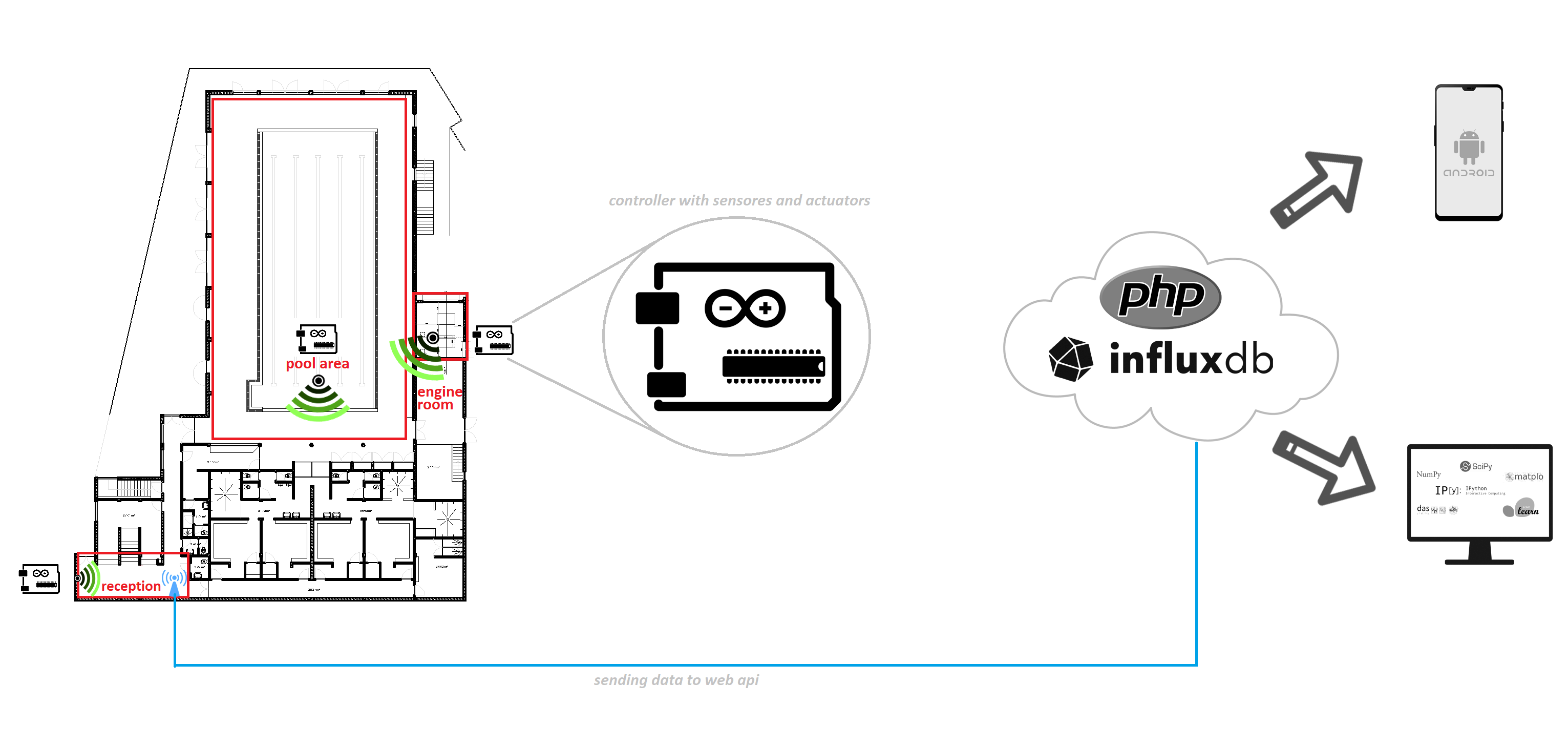}
\caption{\color{Gray} \textbf{Graphical Abstract.}}
\label{figura:graphical_abstract}
\end{figure}

\noindent\textbf{Keywords:} Home Automation, IoT, Arduino, Android, Data
Analysis

% Start line numbers
%\linenumbers

% ---------------------------------------------------------------
\section{Introduction}\label{Introduction}

\begin{figure}[ht]
\centering
\includegraphics[width=0.8\textwidth]{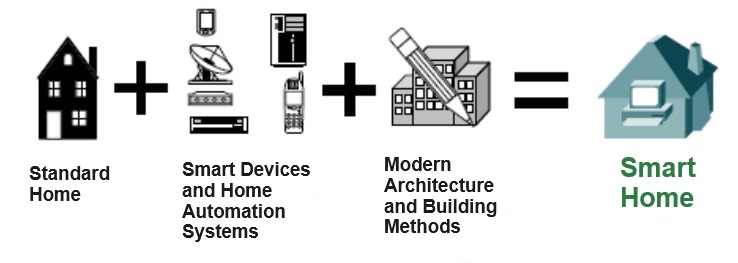}
\caption{\color{Gray} \textbf{Home Automation}~\cite{bolzani2004residencias}.}
\label{figura:homeautomation_1}
\end{figure}

Over the last few years, we have seen a huge increase in the volume of
data, which often generates essential information to aid
decision-making. In the case of the Internet of Things (IoT),
specifically in the area of home automation, there is an increasing
concern to monitor spaces and control equipment remotely via the
Internet, as well as to record the history of data generated by sensors
and actuators connected to devices supported by microcontrollers. Home
automation (Figure~\ref{figura:homeautomation_1}) can be understood as a
network that integrates and controls electronic devices in a home, in
order to meet people's needs and optimise electrical, technological, and
sustainable functions~\cite{shah}.

Smart devices, smart homes, and home automation are terms that are
increasingly being used. The term ``home automation'' comes from the
combination of the words \textit{domus} (house, in Latin) and robotics.
It can be defined as a set of services integrated into a system to
satisfy the basic needs of the occupants of the environment~\cite{bolzani2004residencias}.
Automation is carried out to switch electrical devices on or off by
means of boards incorporating a microcontroller. The actuators,
connected to the boards incorporating microcontrollers, are managed via
a server, which stores a history of events and determines their
operation through coding~\cite{alves2022domotica}.

The municipal swimming pool is widely used by the general population,
from children to senior citizens, and is heated by propane gas. However,
some problems have arisen that have had to be minimised or resolved,
such as the lack of gas leak sensors in the boiler room and the absence
of monitoring of the temperature and humidity of the area surrounding
the pool. We also found that there was no air quality control in the
reception area, which welcomes many users at certain times of the day
during the first week of each month, due to the payment of monthly fees.
Additionally, there have been incidents of burglaries during the night.
All the problems encountered prompted a needs assessment, as well as a
research process and the proposal to monitor some spaces, described in
this paper, so that real-time control could be established to ensure
the safety of users and to record the unplanned entry of people outside
working hours.

This work provides the following contributions:

\begin{itemize}
    \item Design of an effective framework for monitoring a Municipal
          Swimming Pool infrastructure.
    \item Prototype of the proposed framework for three zones (reception,
          pool area, and engine room) inside the public swimming pool
          facilities, including the hardware with connection diagrams and
          a short description of the software for the different layers
          (IoT, middleware, and mobile application).
    \item Analysis of the data collected for decision support.
\end{itemize}

The remainder of this paper is organised as follows.
Section~\ref{Related Work} presents related works on similar problems
and solution proposals by other authors.
Section~\ref{Experimental Work} describes the experimental work, from
the sending of data by microcontrollers to the presentation of this data
in an Android application, including the acquisition of a server and the
development of a Web API. This section also presents the selected
hardware with a brief description.
Section~\ref{Data Analysis and Discussion} presents and discusses the
achieved results. Finally, the conclusions are presented in
Section~\ref{Conclusion}.

% ---------------------------------------------------------------
\section{Related Work}\label{Related Work}

Some literature was found on the potential use of IoT devices for
monitoring physical spaces. Such devices collect and send data to a
cloud platform and, in real time, notify responsible persons to take
appropriate action. The importance of real-time monitoring is mentioned
by Liu~(2016)~\cite{liu2016green} in an article in which he proposed an
air conditioning system in a datacenter, supported by cloud technology,
with the aim of reducing energy consumption without compromising the
operation of the equipment inside this IT infrastructure. The
datacenter's environmental monitoring system, which operates over the
Zigbee protocol~\cite{gupta}, included air conditioning, ventilation,
and temperature control equipment using sensors, whilst the work in the
cloud consisted of storing and collecting data.

Ramphela~(2020)~\cite{ramphela2020internet} defined an integrated
system, also in a datacenter, which includes a database for recording
activities (logs), alerts via email, and a program for controlling
various sensors, which takes responsibility for collecting data on the
monitored variables, such as temperature, humidity, location, movement,
smoke, water, and voltage variation. Similarly to other studies, the aim
is to monitor, in real time, variables that compromise the proper
functioning of a datacenter.

Misra~(2018)~\cite{misra2018iot} describes the use of a system to
monitor a landfill, where sensors (MQ-135 and MQ-136) are used to
detect harmful gases in the space and the maximum waste limit. This
system, which uses an Arduino microcontroller and ESP8266 Wi-Fi module
to connect to the Internet, sends data directly to the municipal
authority, which acts on the values received on a smartphone
application.

Saha~(2017)~\cite{saha2017data} presents temperature monitoring at
various points in a datacenter, using the ESP8266 Wi-Fi microcontroller
and the DHT11 sensor, so that when a certain value is reached, alerts
are sent via SMS and email, in order to identify and solve the problem
in good time, minimising any impact.

Although the study was published some years ago, the layered
architecture of the IoT-based Smart Home System described by
Bing~(2011)~\cite{bing} remains very much relevant. The smart home
system, as proposed in this project, is divided into three layers:
1)~application layer; 2)~network layer; 3)~detection layer. The
detection layer is responsible for collecting data from all household
appliances and sending this data to the intermediate layer, the network
layer. The network layer uses the Internet to send data to the
application layer, which has different functions at different levels
for different purposes. Valente~(2022)~\cite{valente} also emphasises
the use of IoT technology for the correct and controlled use of water
in a vineyard area in Portugal. The use of the LoRaWAN
protocol~\cite{lora} for long-distance data transmission, rather than
Wi-Fi, is especially important for applications in remote areas where
GSM networks have poor coverage. The data collected by sophisticated
sensors is stored in a specific time-series database
(InfluxDB)~\cite{influxdb}, and a predictive machine learning model is
applied to determine and predict water stress, among other factors.

Another investigation~\cite{flores-cortez} involved monitoring air
quality in the city of El Salvador. In this case, an ESP32 controller
was used together with a Wemos-Lolin development board, to which a
PMS5003 particle contamination sensor was connected. On the software
side for the IoT platform, a storage, graphics, and website system was
implemented based on low-cost tools. The main result obtained in this
work was an IoT prototype of an electronic station that makes it
possible to monitor levels of contamination by particulate material in
the environment, with the data accessible from any device with internet
access via a website.

Gresa~(2021)~\cite{SanzGresa2021} raised the issue of pool maintenance
concerns and the positive effects of using IoT technology to reduce
costs. The proposed solution is based on a web application, developed in
Hypertext Preprocessor (PHP), which receives data, also in real time,
from an Arduino microcontroller, to control the water level, ambient
temperature, and atmospheric pressure.

Another study~\cite{rocha2023}, which serves as the basis for this
project proposal, focuses on a low-cost solution for monitoring a
datacenter as an information technology infrastructure for disaster
recovery, using only an Arduino Uno microcontroller, sensors for
temperature, humidity, light, and movement, and a magnetic sensor (the
latter to check for a breach of the space -- an unwanted or unplanned
entry into the area). The data is stored in a MySQL database and
displayed in an Android application, where there is also the possibility
of receiving alerts via email, programmed in an API developed in PHP.

% ---------------------------------------------------------------
\section{Experimental Work}\label{Experimental Work}

Home automation has been widely adopted as an efficient solution to
improve the safety, comfort, and energy efficiency of environments.
With the advancement of technologies and the emergence of
interconnected devices, such as sensors, actuators, and microcontrollers
coupled to Arduino boards, it became possible to develop intelligent
systems capable of controlling and monitoring various aspects of a
residence or, in this case, a public service infrastructure. The object
of study, as mentioned above, is the Vouzela Municipal Swimming Pool.
This section describes the steps taken to assemble the hardware in three
selected spaces (Reception, the area around the pool/tank, and the
Engine Room, where the boiler is located), as well as a brief
description of this hardware, including a connection diagram for each
space.

The Arduino Mega board was installed in the Reception area, where four
sensors are connected to monitor and control different aspects of the
environment, whilst two Arduino Uno boards were installed in the Engine
Room area in order to monitor the boiler space due to possible gas
leaks, as well as the pool area (tank) for environmental control. It is
worth noting that each Arduino board required connection to another
microcontroller (ESP8266-01S) in order to send the data to the
Internet.

% ---------------------------------------------------------------
\subsection{Brief Hardware Description}

\begin{figure}[ht]
\centering
\includegraphics[width=0.7\textwidth]{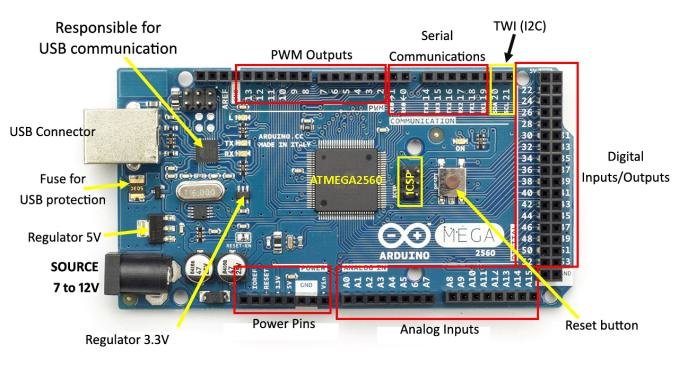}
\caption{\color{Gray} \textbf{Arduino Mega}~\cite{mega}.}
\label{figura:arduinomega}
\end{figure}

The Arduino Mega (Figure~\ref{figura:arduinomega}), equipped with the
ATmega2560 microcontroller, stands out for its expanded input and output
capacity, making it ideal for projects that require a greater number of
connections. The ATmega2560 microcontroller, present in the Arduino
Mega, has an 8-bit RISC architecture, an operating frequency of 16~MHz,
and a program memory of 256~KB. This expanded memory capacity is
especially useful for more complex projects that require storing a
greater volume of code and data.

\begin{figure}[h]
\centering
\includegraphics[width=0.7\textwidth]{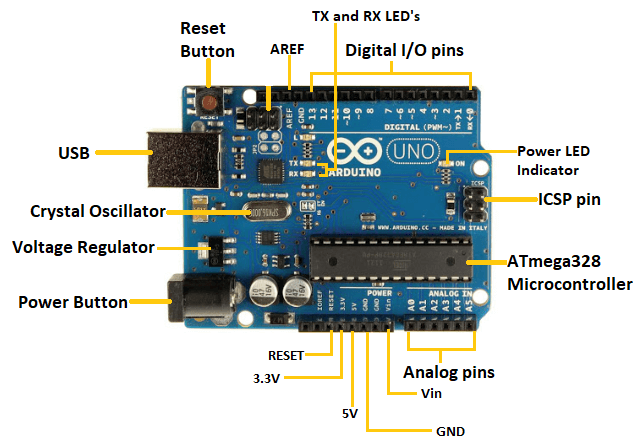}
\caption{\color{Gray} \textbf{Arduino Uno}~\cite{uno}.}
\label{figura:arduinouno}
\end{figure}

The Arduino Uno (Figure~\ref{figura:arduinouno}) has an ATmega328P
microcontroller and is characterised by its simplicity and ease of use.
This board has a sufficient number of input and output pins, making it
suitable for smaller-scale projects. The ATmega328P microcontroller,
used in the Arduino Uno, also features an 8-bit RISC architecture, an
operating frequency of 16~MHz, but a program memory of only 32~KB,
which may not be suitable for more complex projects.

However, for this work, due to the number of hardware components that
naturally consume energy, two Arduino boards were used with independent
power supplies, so that the integrity of the data being constantly
collected was not compromised. Both Arduino boards include digital
input/output pins, a USB connection, a power connector, and a reset
button. The Arduino is an open-source product (board) integrating a
processor and a bootloader, ready to be connected via USB to a computer.

\begin{figure}[h]
  \begin{subfigure}{.2\columnwidth}
    \centering
    \includegraphics[width=\textwidth]{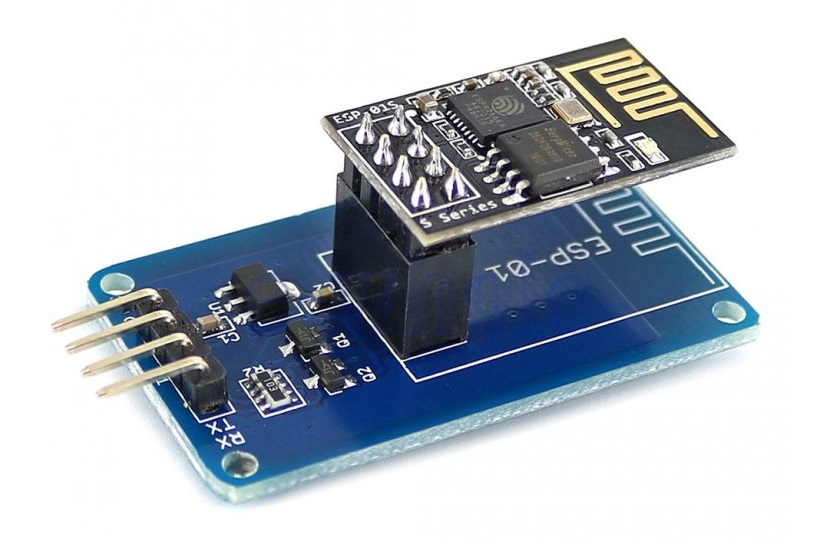}
    \caption{}
    \label{fig:esp8266_wifi_module}
  \end{subfigure}
  \begin{subfigure}{.2\columnwidth}
    \centering
    \includegraphics[width=\textwidth]{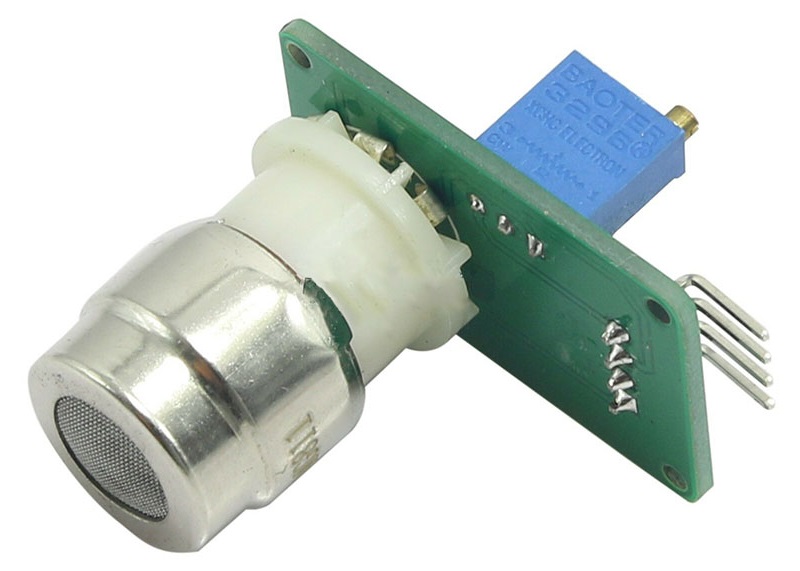}
    \caption{}
    \label{fig:CO2_sensor}
  \end{subfigure}
  \begin{subfigure}{.2\columnwidth}
    \centering
    \includegraphics[width=0.9\textwidth]{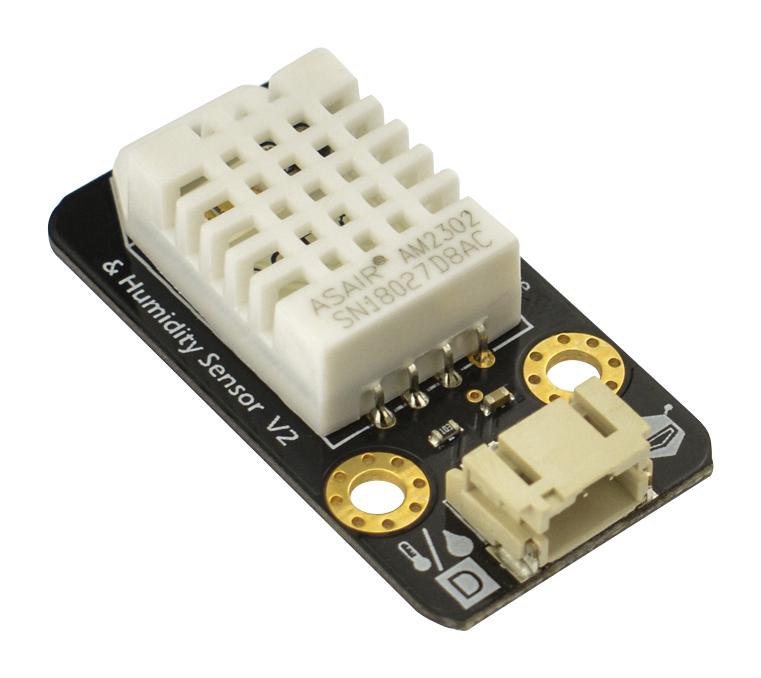}
    \caption{}
    \label{fig:dht22}
  \end{subfigure}
  \begin{subfigure}{.2\columnwidth}
    \centering
    \includegraphics[width=0.8\textwidth]{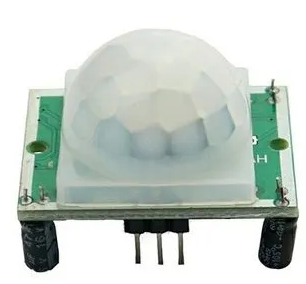}
    \caption{}
    \label{fig:pir}
  \end{subfigure}
  \begin{subfigure}{.2\columnwidth}
    \centering
    \includegraphics[width=0.8\textwidth]{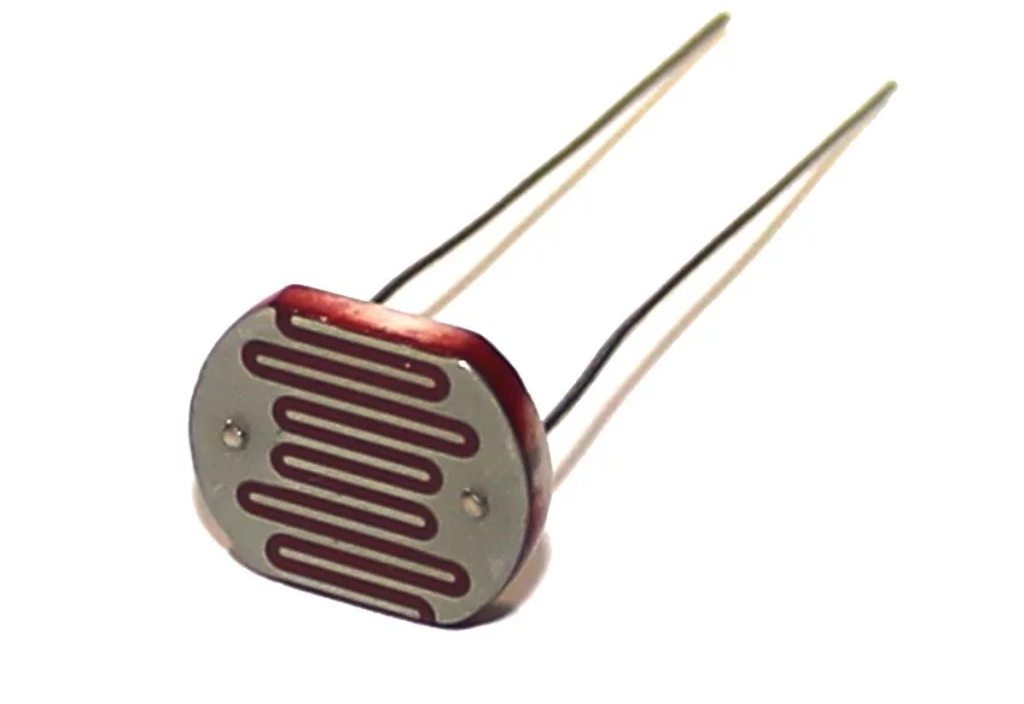}
    \caption{}
    \label{fig:ldr}
  \end{subfigure}
  \begin{subfigure}{.2\columnwidth}
    \centering
    \includegraphics[width=\textwidth]{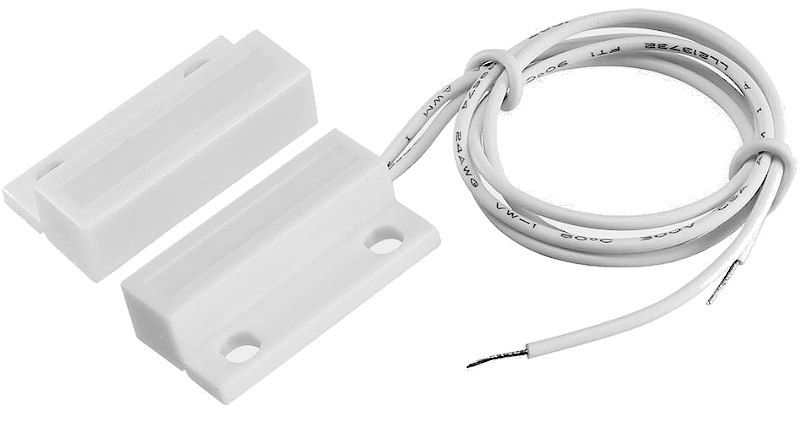}
    \caption{}
    \label{fig:magnetic}
  \end{subfigure}
  \begin{subfigure}{.2\columnwidth}
    \centering
    \includegraphics[width=\textwidth]{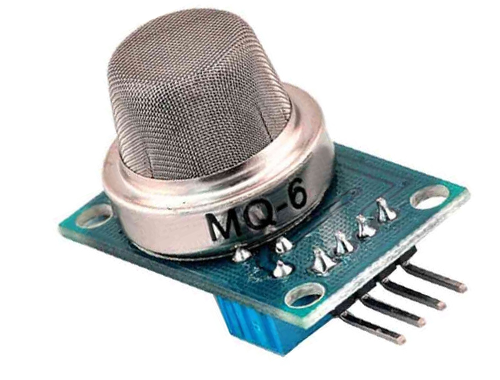}
    \caption{}
    \label{fig:gassensor}
  \end{subfigure}
  \begin{subfigure}{.2\columnwidth}
    \centering
    \includegraphics[width=\textwidth]{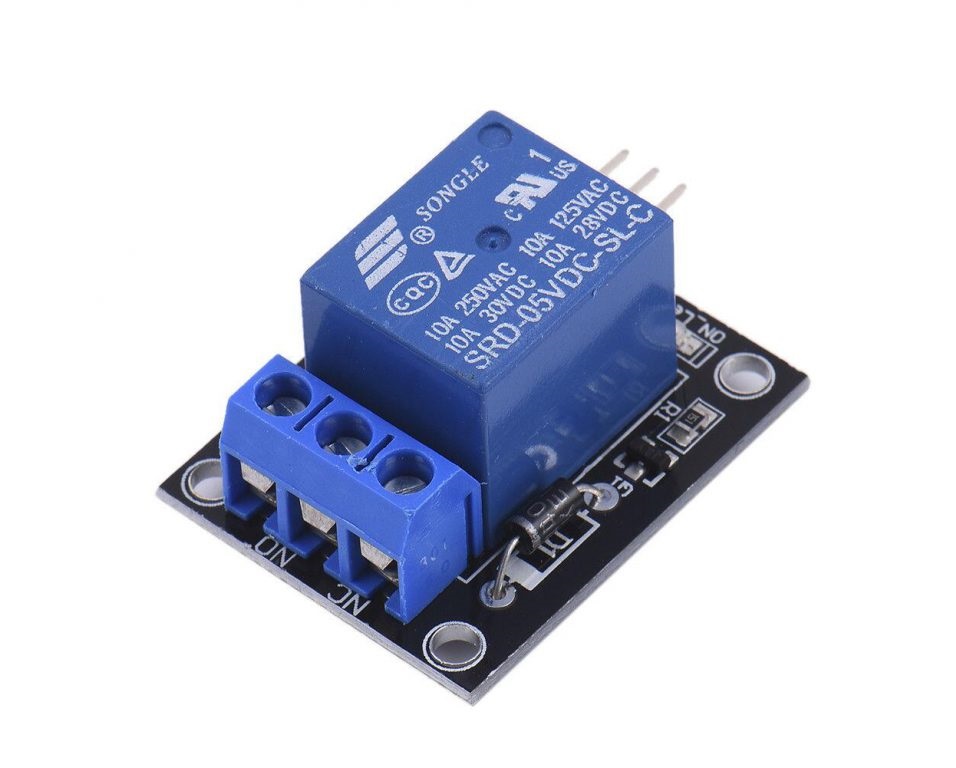}
    \caption{}
    \label{fig:rele}
  \end{subfigure}
  \begin{subfigure}{.2\columnwidth}
    \centering
    \includegraphics[width=\textwidth]{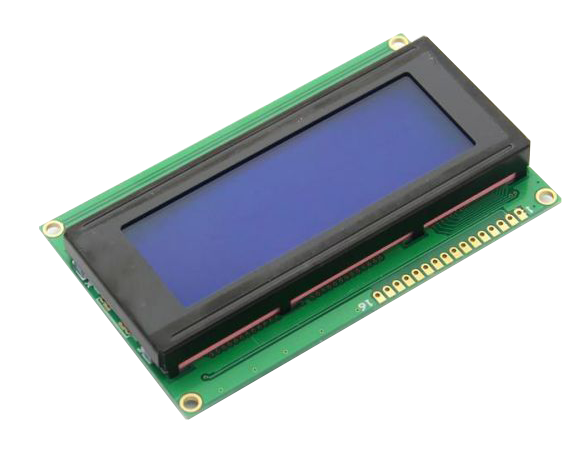}
    \caption{}
    \label{fig:lcd}
  \end{subfigure}
  \caption{\color{Gray} \textbf{ESP microcontroller, sensors, and
    actuators.}
    (a)~ESP8266 Wi-Fi module;
    (b)~CO2 sensor;
    (c)~temperature and humidity sensor;
    (d)~motion sensor;
    (e)~light sensor;
    (f)~magnetic door sensor;
    (g)~propane gas sensor;
    (h)~relay actuator;
    (i)~LCD actuator.}
  \label{fig:IoT_Sensors}
\end{figure}

The ESP8266-01S wireless module, as illustrated in
Figure~\ref{fig:esp8266_wifi_module}, allows the connection of a
microcontroller to a Wi-Fi network, efficiently and at reduced cost. It
supports 802.11~b/g/n networks and can operate as an Access Point or as
a Station, sending and receiving data~\cite{satapathy2018arduino} using
AT commands~\cite{esp01}. In practice, the ESP8266 Wi-Fi module
connects to wireless data networks to exchange information between
microcontrollers and, as in this project, to a web API on the
Internet~\cite{esp01toapi}. The module's communication with the Arduino
can be accomplished via serial using the RX and TX pins, and can be
configured through AT commands. We recommend the easy-to-use Wi-Fi
adapter ESP-01, which already has the converter and can be connected
directly to the breadboard.

For reasons of health and well-being of users, the MG-811 sensor
(Figure~\ref{fig:CO2_sensor}) is essential in the automation system.
This sensor is responsible for measuring carbon dioxide (CO2) levels
in the environment. However, these values may vary depending on the
temperature of the location under study~\cite{Saptiani_2019}. If there
is a high level of CO2, indicating poor air quality, it sends a signal
to the Arduino Mega, which activates an actuator (relay) to turn on a
fan, for example. This action aims to recirculate the air and ensure a
cleaner and healthier environment.

The digital humidity and temperature sensor (DHT22), as presented in
Figure~\ref{fig:dht22}, measures temperature in the range of
$[-40,\,80]\,^{\circ}$C and air humidity in the range of 0 to 100\%,
with an accuracy of approximately 2\%~\cite{dht22}. This DHT22 sensor
detects humidity and temperature and sends this information to the
microcontroller board, which must be programmed to perform an action
when a certain humidity or temperature value is reached. The DHT22
sensor comprises a capacitive humidity sensor and a thermistor to
measure the surrounding air, and sends a digital signal to the data
pin.

The passive infrared (PIR) sensor, shown in Figure~\ref{fig:pir}, can
detect objects moving within an area of a few metres. Thus, it can be
used to trigger a notification when someone moves into this area. It is
important to note that this sensor only detects movement (heat change)
and not presence; that is, if something remains stationary in front of
it, the PIR sensor does not detect it~\cite{pir}. The PIR sensor for
Arduino has another characteristic, which is the delay demonstrated
right after detecting an object. This is because, on its back, the PIR
sensor has a potentiometer that allows a minimum delay setting of
approximately 3 seconds~\cite{pirspecs}.

The Light Dependent Resistor (LDR), presented in Figure~\ref{fig:ldr},
is a component whose resistance varies according to light intensity. The
LDR, or photoresistor, is a type of resistor that is variable in
nature. Its resistance changes based on the intensity of light it
receives. When exposed to more light, the LDR's resistance decreases,
and when subjected to less light, its resistance increases. It is
important to note that, similar to a standard resistor, the LDR is
non-polarised, and its resistance is quantified in ohms. In darkness,
its resistance is approximately 1~M$\Omega$, whilst in the presence of
light, it typically ranges between 10 and 20~k$\Omega$~\cite{ldr}.
This variation is used, in this work, to indicate whether the light is
on or off; that is, after several tests, if the value is very low, the
light can be confirmed as being off.

A set of magnetic sensors, as depicted in Figure~\ref{fig:magnetic},
is designed to notify operators when doors or windows open. These
sensors are mainly used in residential or industrial security
systems~\cite{mc38}.

The MQ-6 sensors, as depicted in Figure~\ref{fig:gassensor}, use a
metal oxide material which has electrical conductivity properties that
vary in response to the presence of specific gases~\cite{ajiboye}.
Specifically, for this work, two sensors were used to detect propane
gas in the boiler area.

With regard to the actuators, two relays, as shown in
Figure~\ref{fig:rele}, were used: one for the Reception area and the
other for the boiler area, in order to act when the CO2 value in the
reception space reaches a predefined value, and when the propane gas
value in the boiler area reaches a value programmed in the
microcontroller. The 5~V 1-Channel Relay Module allows the simple and
practical control of AC (alternating current) loads from a
microcontrolled platform. With only 1 channel, it can control a single
AC load of up to 10~A. It is commonly used in home automation projects
for controlling lights, fans, and other outputs that can be activated
through a relay~\cite{rele}.

In addition to the sensors and actuators mentioned above, it is
important to note that each Arduino board is connected to a Liquid
Crystal Display (LCD), which plays a crucial role in displaying
information in real time. These LCDs (Figure~\ref{fig:lcd}) provide
visual feedback to users, displaying relevant data on the values
collected and processed by the microcontrollers~\cite{lcd}, without
having to consult the Android application. An interesting feature of
these screens is the ability to intelligently control the backlight.
Using the PIR sensors previously installed, it is possible to detect
movement next to the boxes where the Arduinos and breadboards are
installed. When a person in charge approaches, the PIR sensors detect
movement and activate the backlight of the LCD screens, making
information visible in a darker environment. This functionality saves
energy by turning off the backlight when not needed, and provides an
intuitive and convenient experience for those wishing to view
information.

% ---------------------------------------------------------------
\subsection{Hardware Assembly and Data Presentation}

\begin{figure}[ht]
\centering
\includegraphics[width=0.7\textwidth]{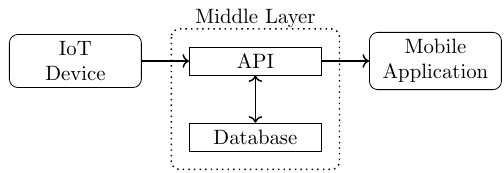}
\caption{\color{Gray} \textbf{Proposed framework.}}
\label{figura:Proposed_framework}
\end{figure}

The general architecture of the developed project is depicted in
Figure~\ref{figura:Proposed_framework}, showing its modules: the IoT
devices collect data from the physical spaces of the Municipal Pool;
the middle layer, with the Web API, intermediates communication with
the InfluxDB database and provides data to the Android application.

All hardware was programmed using the C programming language, through
the Arduino IDE. The data collected from sensors are recorded in real
time into a middle-layer server (Linux Ubuntu) providing webserver and
database services. To add some security to the server, the Fail2Ban
service was also installed, whose objective is to temporarily block
unauthorised or unknown access attempts~\cite{fail2bangithub}. An API
developed in PHP and deployed on an Apache webserver provides services
receiving data from the IoT devices (Arduino microcontrollers). The
data was stored in a time-series database
(InfluxDB)~\cite{influx1,influxphp} and subsequently sent to the Android
application. The choice regarding data storage fell on the InfluxDB
platform, as it appears to be the solution for storing a large amount of
data generated by IoT applications, where superior performance is
demonstrated compared to other more conventional database
engines~\cite{naqvi2017time}.

Moreover, the API makes the collected data available from the database
to a mobile application developed in Java for Android.

The web application relies on the PHPMailer~\cite{phpmailer} library to
notify operators via email, through the use of the SparkPost
service~\cite{sparkpostr}. In addition to notifications via email, the
person responsible for the municipal public service is also able to
receive push notifications through the OneSignal
service~\cite{onesignaldoc,onesignalgithub}. These push notification
settings can be configured in the Android application. As mentioned, the
proposed framework targets three areas in the municipal swimming pool
infrastructure. In what follows, we describe the work carried out in
each of these spaces: Reception, Swimming Pool (tank) area, and Engine
Room. In summary, the data is sent from the Arduinos to the server
through the WifiEsp library~\cite{wifiesp}. The Web API processes the
data and stores it in a time-series database. When the Android
application makes a request for stored data through the OkHttp
library~\cite{okHttp}, the Web API returns the data in JSON format and
the Android application interprets this data through the Gson
library~\cite{gson} and displays it in the indicated interfaces.

% ---------------------------------------------------------------
\subsubsection{Reception Area}

\begin{figure}[ht]
\centering
\includegraphics[width=0.8\textwidth]{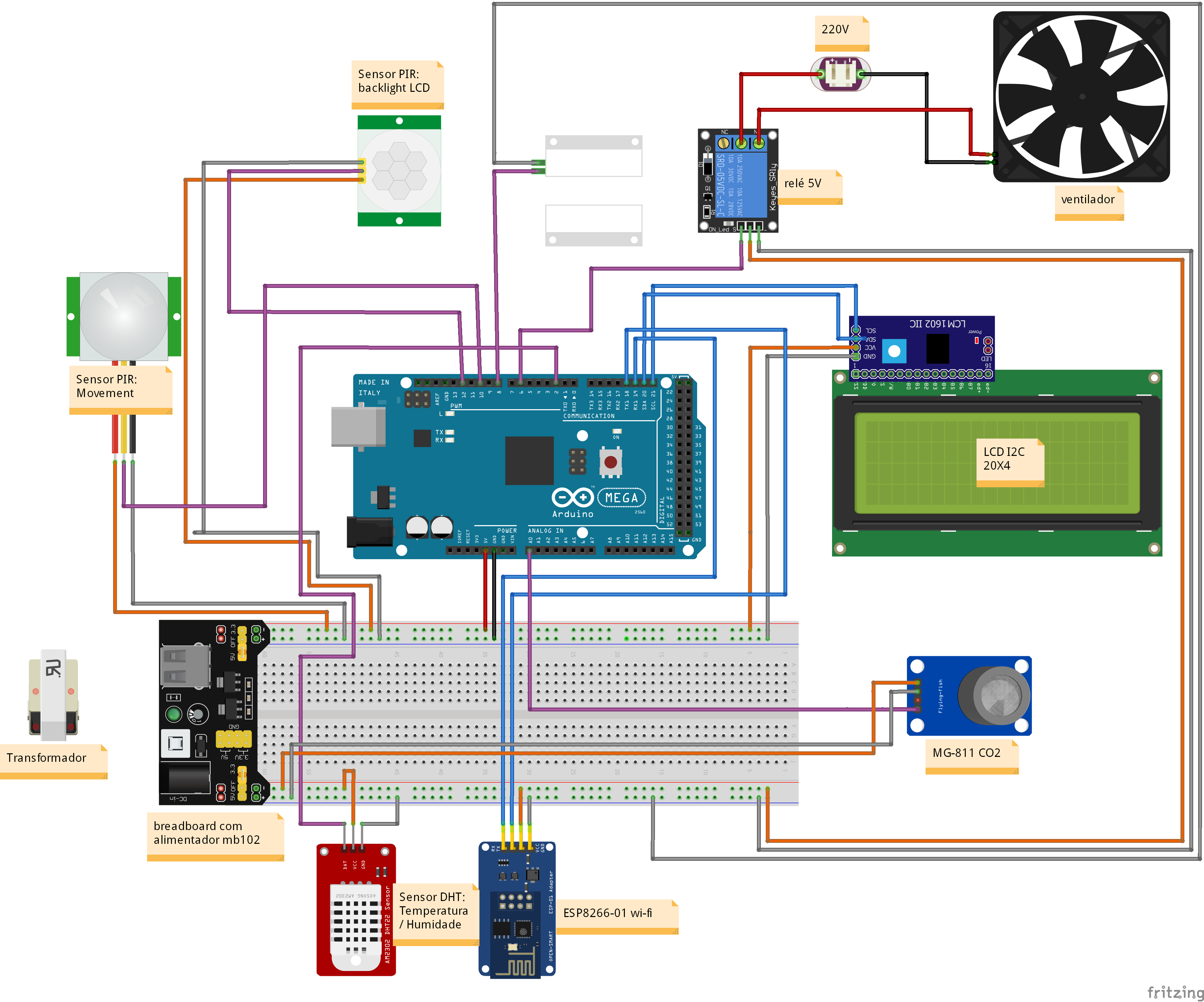}
\caption{\color{Gray} \textbf{Reception area -- hardware connection
    diagram.}}
\label{figura:Reception_area_connection_diagram}
\end{figure}

In the Reception area of the Vouzela municipal swimming pool, the
Arduino Mega plays a crucial role in the home automation system. As
already mentioned, this board, known for its wide data input and output
capacity, allows the connection and control of various devices, including
sensors and actuators. Figure~\ref{figura:Reception_area_connection_diagram}
shows the connection diagram for all sensors and actuators connected to
the Arduino Mega.

\begin{figure}[!h]
  \centering
  \begin{subfigure}{.4\textwidth}
    \centering
    \includegraphics[width=\textwidth]{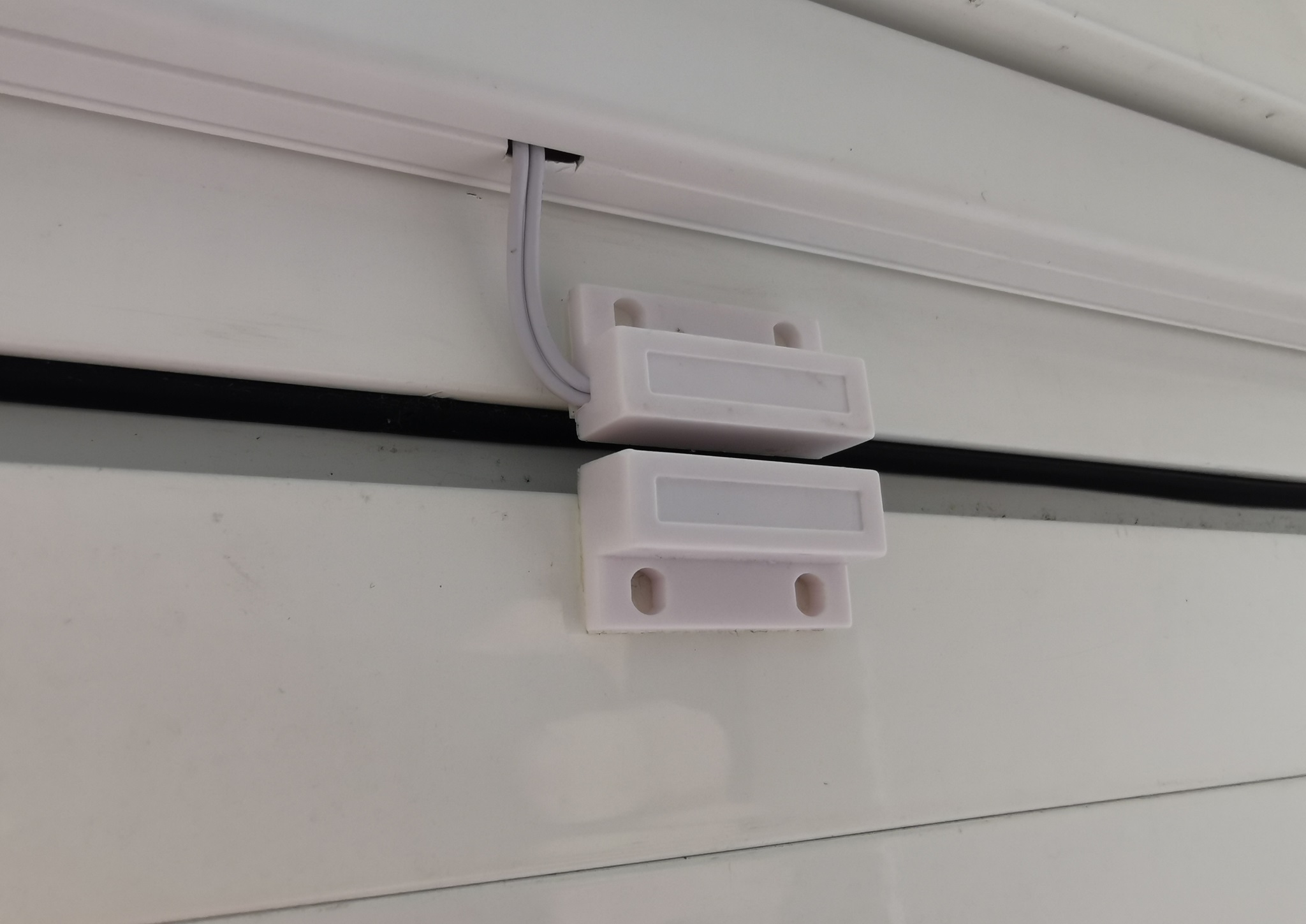}
    \caption{}
    \label{figura:Magnetic_sensor_door}
  \end{subfigure}%
  \qquad
  \begin{subfigure}{.4\textwidth}
    \centering
    \includegraphics[width=\textwidth]{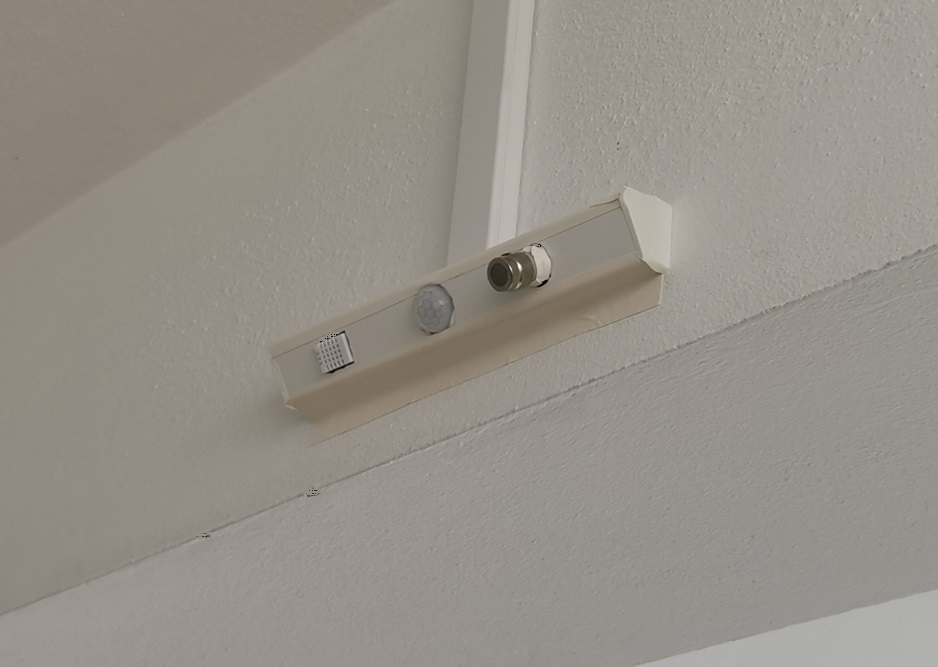}
    \caption{}
    \label{figura:DHT_PIR_CO2_reception}
  \end{subfigure}
  \caption{\color{Gray} \textbf{Sensors in the Reception area.}
    (a)~Magnetic sensor on the door;
    (b)~temperature and humidity, motion, and CO2 level sensors.}
  \label{fig:all_sensors_reception}
\end{figure}

The microcontroller coupled to the Arduino Mega is responsible for
collecting the data read by the sensors and sending it to the web API
through another microcontroller, the ESP8266-01. The parameters
analysed in the Reception area include environmental temperature and
humidity, movement, door opening, and, for the health of users and
employees, carbon dioxide (CO2) levels. A real image of the sensors
installed in the Reception area is shown in
Figure~\ref{fig:all_sensors_reception}.

\begin{figure}[ht]
\centering
\includegraphics[width=0.8\textwidth]{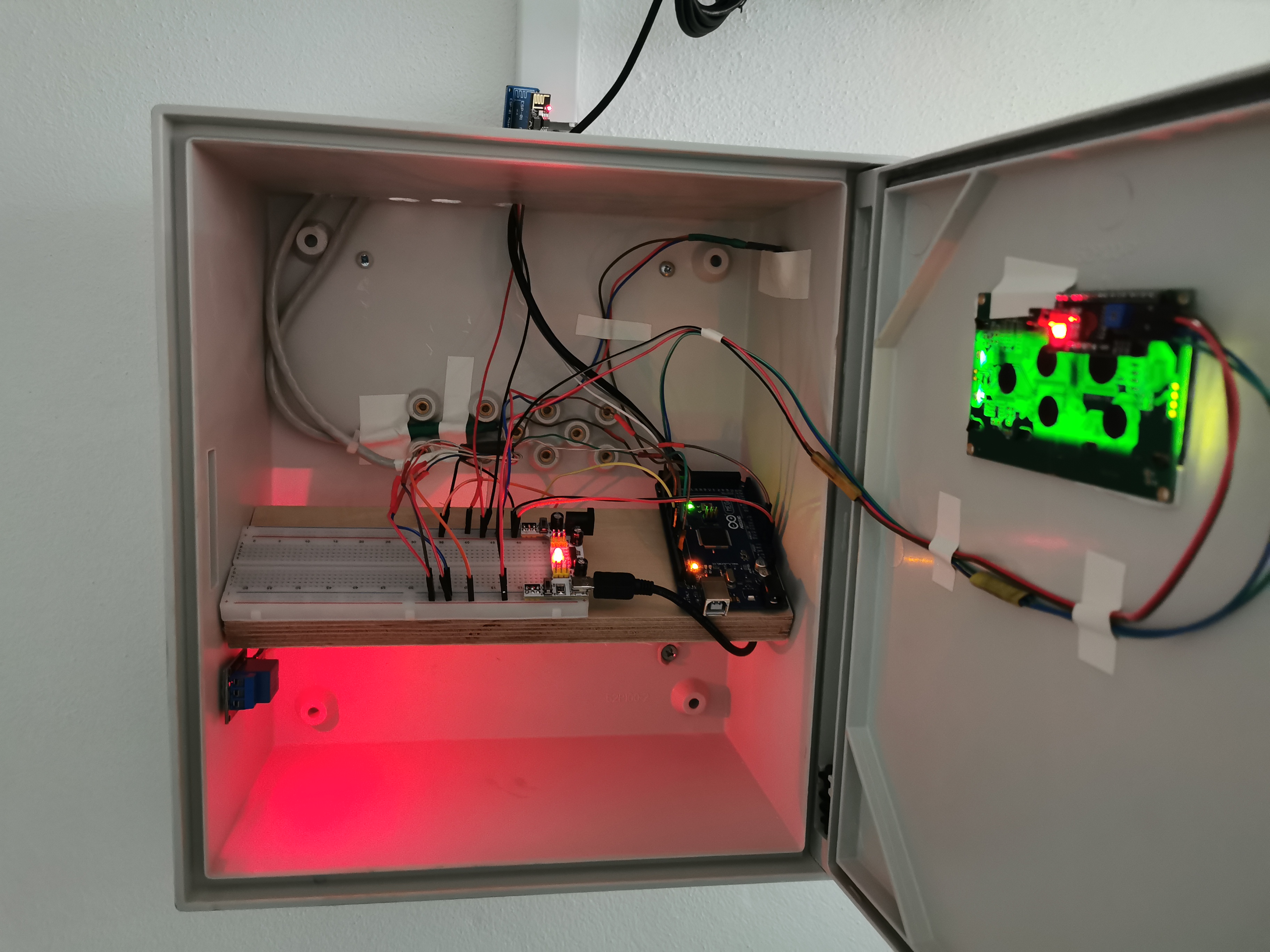}
\caption{\color{Gray} \textbf{Reception area Arduino box.}}
\label{figura:Reception_area_Arduino_Box}
\end{figure}

\begin{figure}[ht]
  \centering
  \begin{subfigure}{.4\textwidth}
    \centering
    \includegraphics[width=\textwidth]{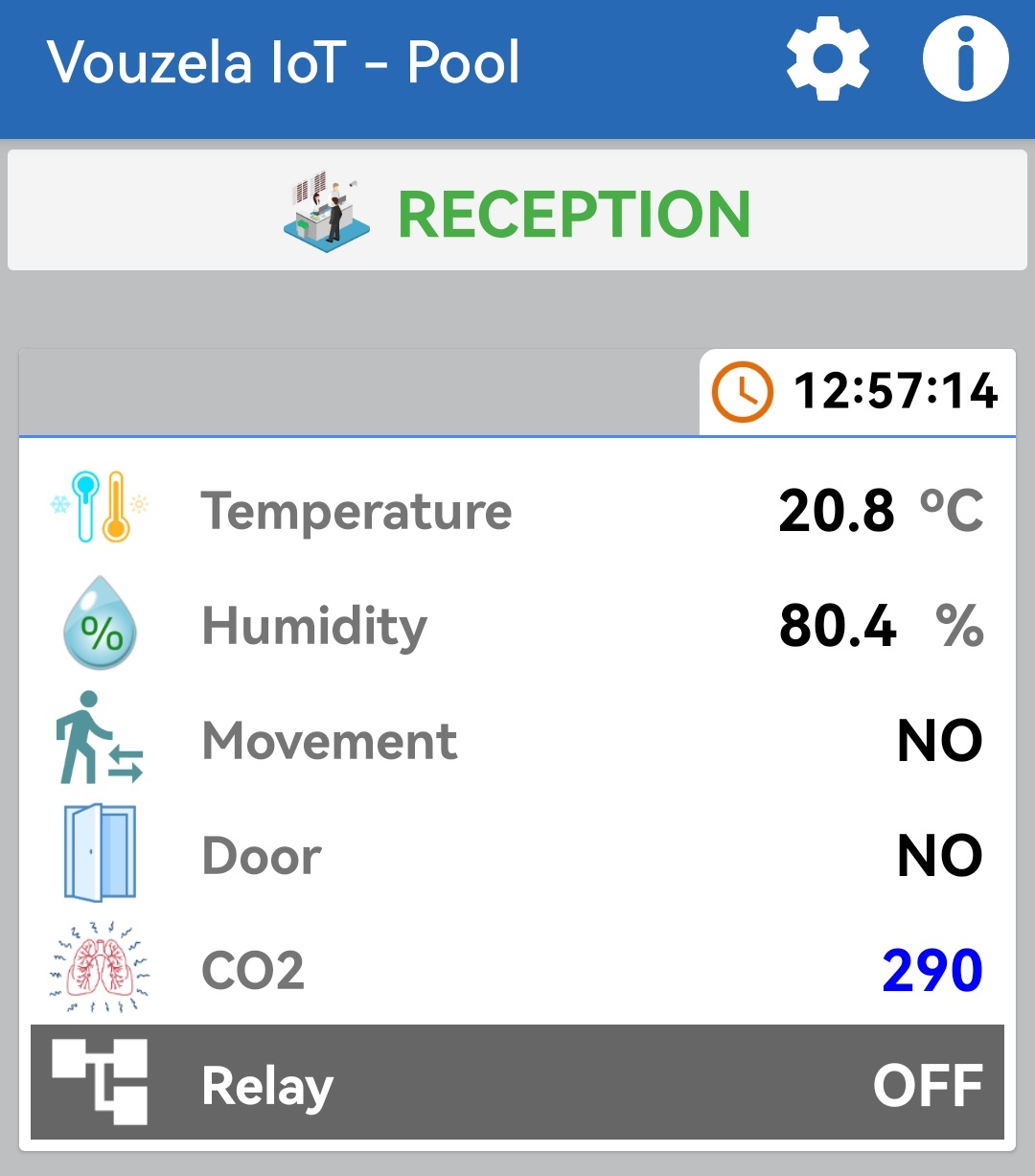}
    \caption{}
    \label{figura:Android_All_Parameters_Reception}
  \end{subfigure}%
  \qquad
  \begin{subfigure}{.4\textwidth}
    \centering
    \includegraphics[width=\textwidth]{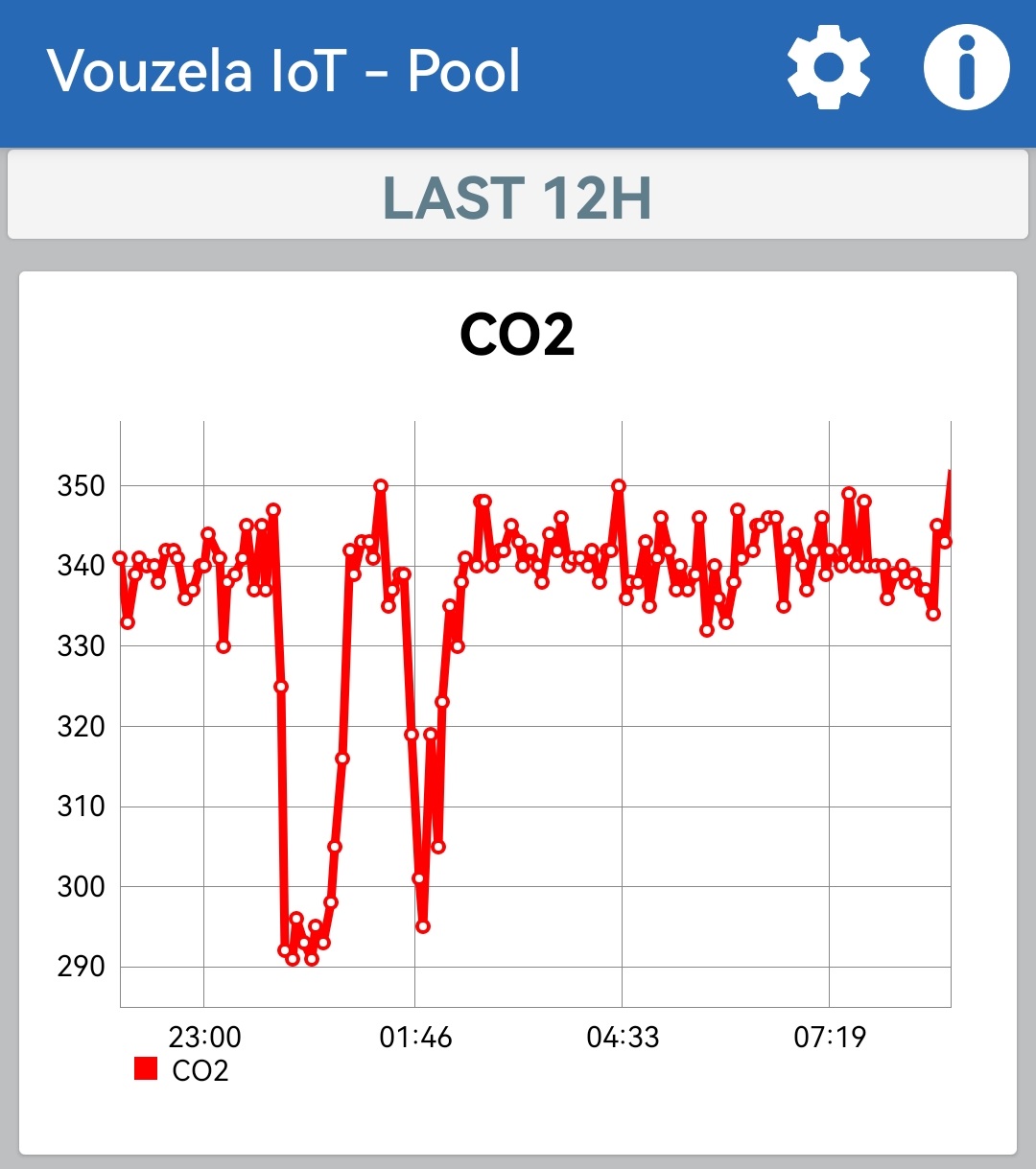}
    \caption{}
    \label{figura:Android_CO2_Graph_Reception}
  \end{subfigure}
  \qquad
  \begin{subfigure}{.3\textwidth}
    \centering
    \includegraphics[width=\textwidth]{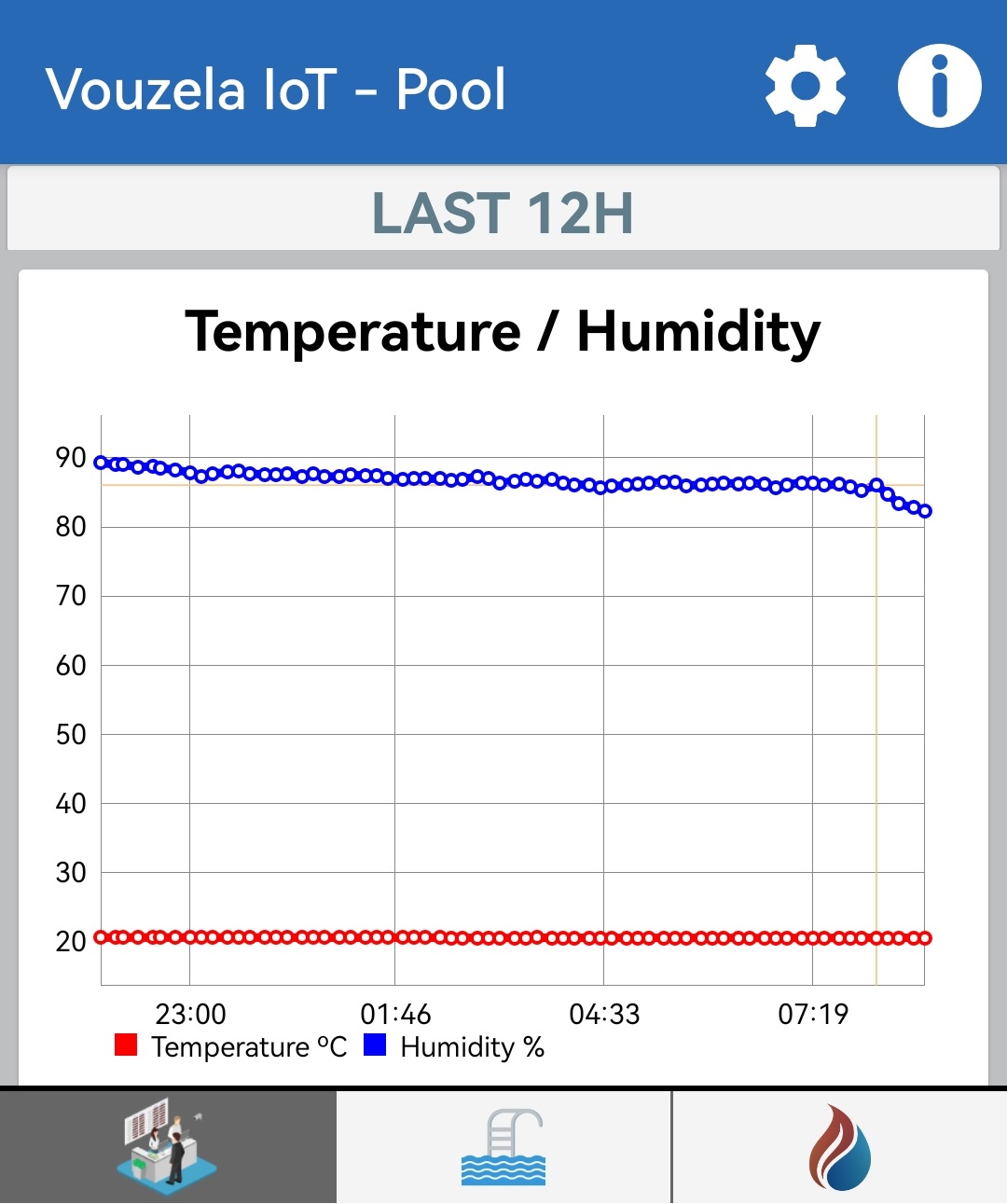}
    \caption{}
    \label{figura:Android_TempHum_Graph_Reception}
  \end{subfigure}\qquad
  \begin{subfigure}{.5\textwidth}
    \centering
    \includegraphics[width=\textwidth]{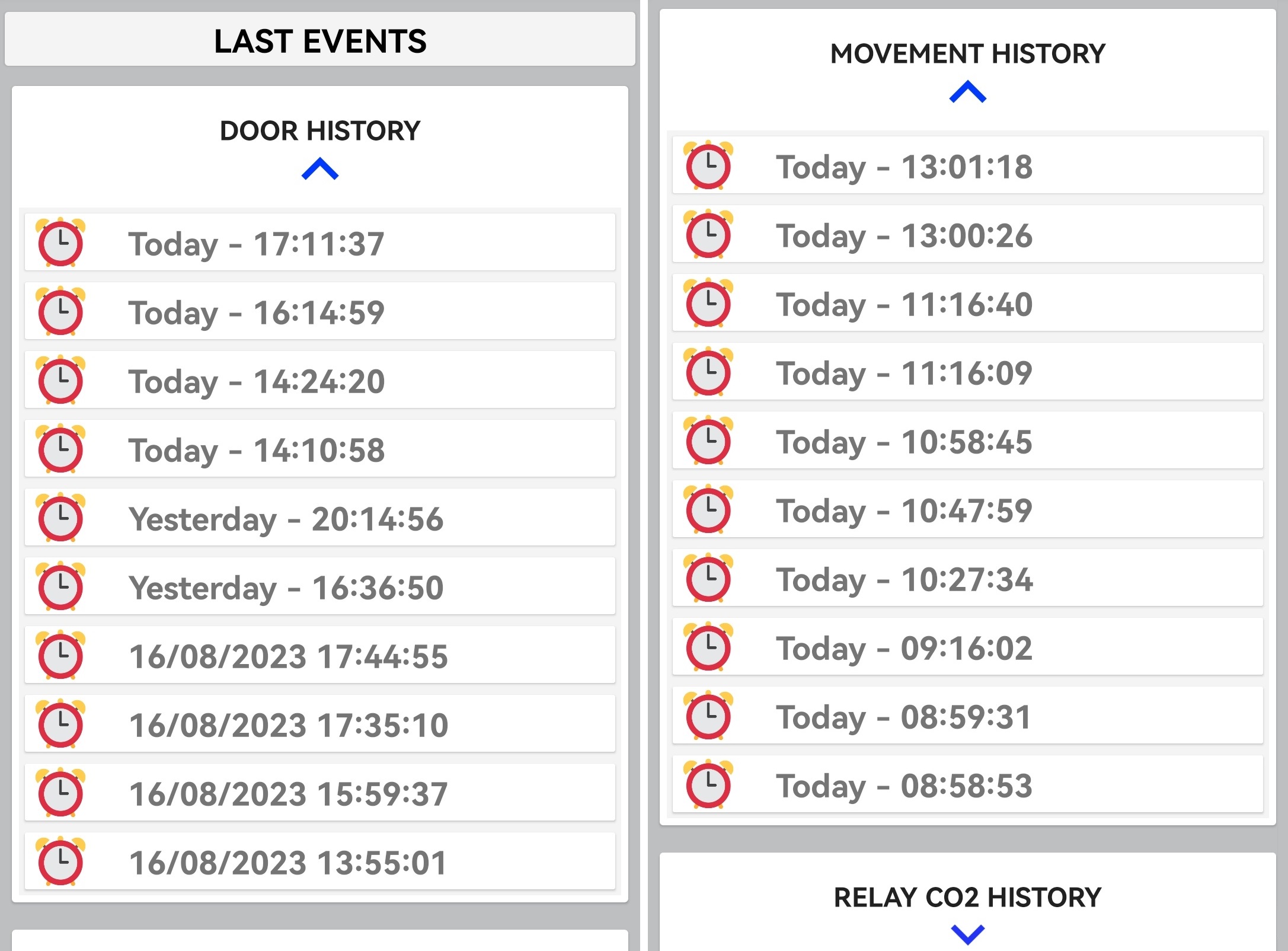}
    \caption{}
    \label{figura:Android_Door_Movement_Reception}
  \end{subfigure}
  \caption{\color{Gray} \textbf{Android mobile application for the
    Reception area.}
    (a)~All parameters in real time;
    (b)~CO2 variation graph over the last 12 hours;
    (c)~temperature and humidity variation graph over the last 12 hours;
    (d)~last door and movement events.}
  \label{fig:android_sensors_reception}
\end{figure}

If the CO2 values reach a predefined value in the microcontroller, a
relay is automatically activated, which can turn on a fan or an air
recirculation device. In Figure~\ref{figura:Reception_area_Arduino_Box}
we can see the box containing a breadboard with an independent power
supply, which provides energy to the microcontrollers (ATmega and ESP),
sensors, and actuators (relay and LCD screen). At the same time, the
values are also displayed on an LCD screen located on the box door. In
addition to the data displayed on the LCD screen in real time, the data
is also sent to the Android application in order to inform those
responsible for this public service. The mobile application shows all
data (Figure~\ref{figura:Android_All_Parameters_Reception}) in real
time, relating to the previously mentioned parameters, as well as a
graph of the variation in values over the last 12 hours in relation to
CO2 levels (Figure~\ref{figura:Android_CO2_Graph_Reception}) and
temperature and humidity
(Figure~\ref{figura:Android_TempHum_Graph_Reception}). Still regarding
the Reception area, the mobile application presents, in the form of a
list, the last 10 records of door opening and movement detected
(Figure~\ref{figura:Android_Door_Movement_Reception}). It should also
be noted that, if the relay connected to the CO2 parameter is
activated, the date and time of each activation of that actuator is
also recorded. Some parameter values will be explored in the following
section, in order to address questions raised, such as the acquisition
of air conditioning equipment and air recirculation equipment, to
maintain the comfort and well-being of users and employees of the
municipal swimming pool.

% ---------------------------------------------------------------
\subsubsection{Swimming Pool (Tank) Area}

\begin{figure}[ht]
  \centering
  \begin{subfigure}{.4\textwidth}
    \centering
    \includegraphics[width=\textwidth]{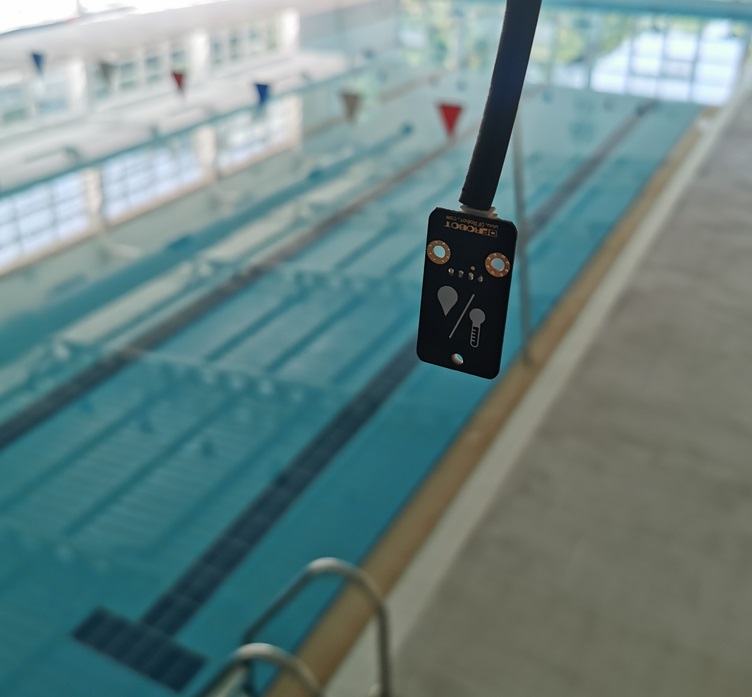}
    \caption{}
    \label{figura:DHT_Pool_area}
  \end{subfigure}%
  \qquad
  \begin{subfigure}{.5\textwidth}
    \centering
    \includegraphics[width=\textwidth]{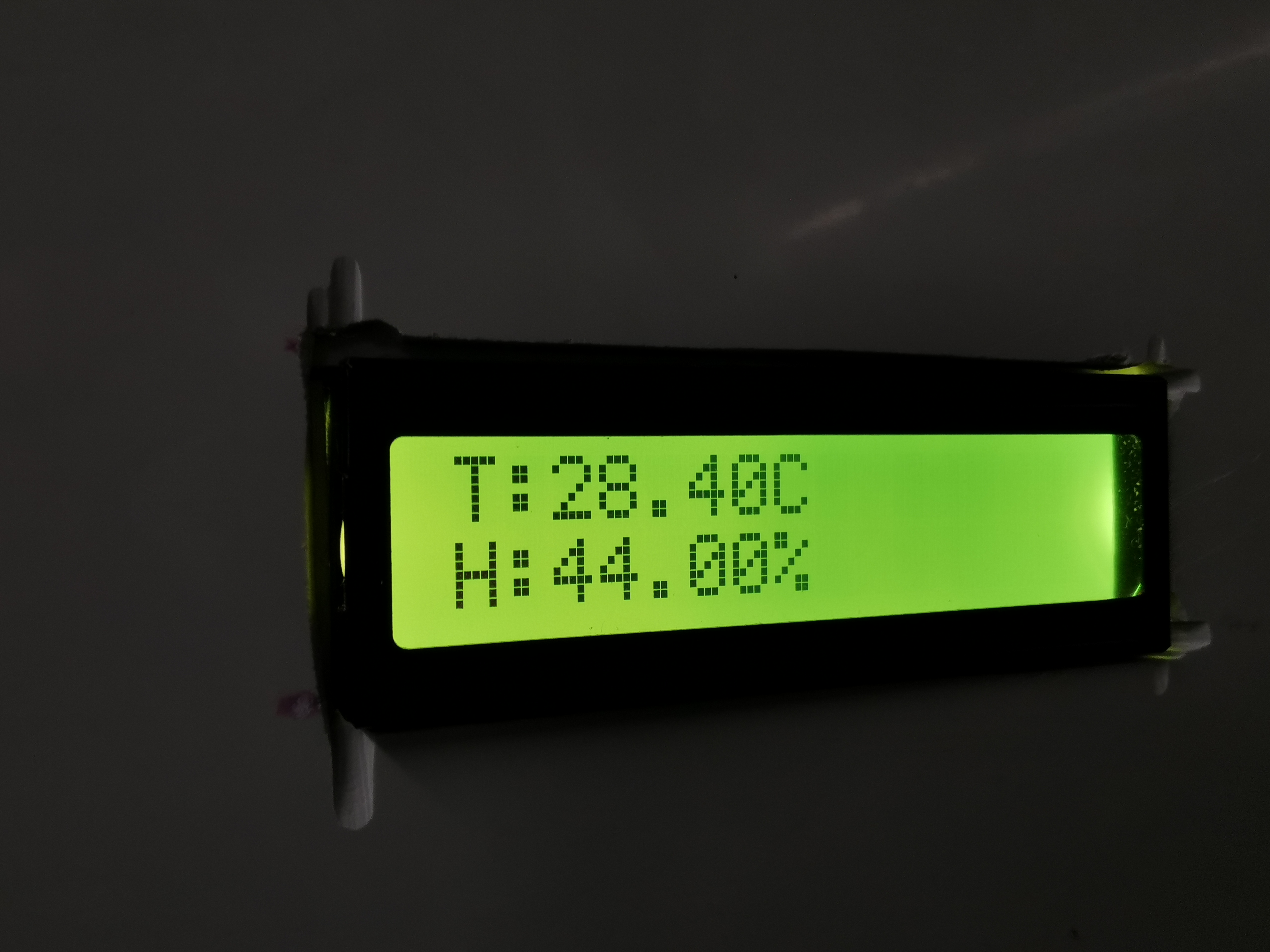}
    \caption{}
    \label{figura:LCD_pool_area}
  \end{subfigure}
  \caption{\color{Gray} \textbf{Sensors connected to the Arduino
    responsible for sending pool environment values.}
    (a)~DHT sensor in the space surrounding the water tank;
    (b)~real-time temperature and relative humidity on the LCD screen.}
  \label{fig:all_sensors_pool}
\end{figure}

\begin{figure}[h]
\centering
\includegraphics[width=.7\textwidth]{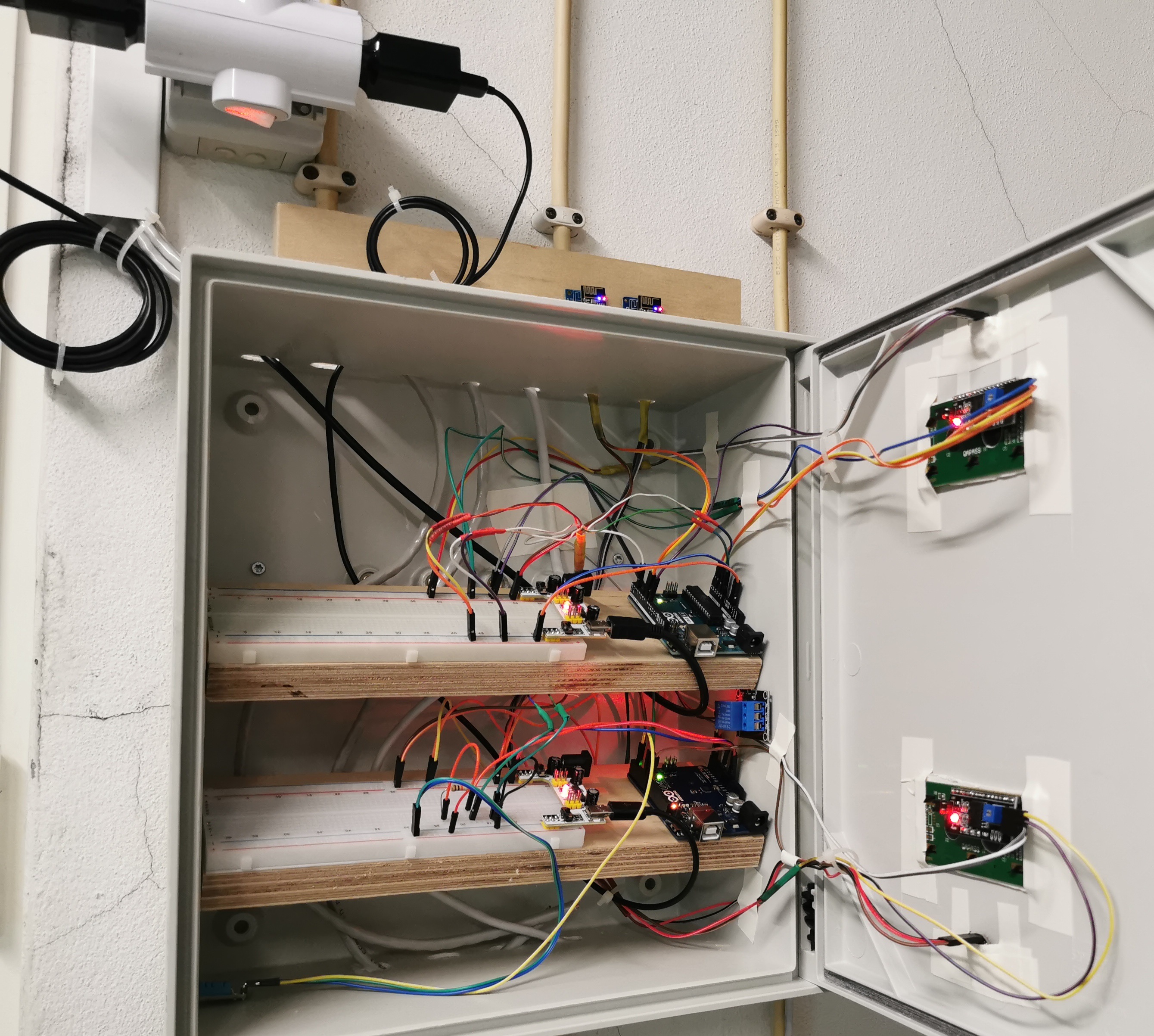}
\caption{\color{Gray} \textbf{Engine room Arduino box.}}
\label{figura:Engine_room_Arduino_Box}
\end{figure}

Like the Reception area, the temperature and humidity in the space
surrounding the water tank (swimming pool) was also subject to constant
monitoring. A temperature and humidity sensor (DHT22) was installed in
a strategic location, not easily visible and difficult for users to
access (Figure~\ref{figura:DHT_Pool_area}). This DHT sensor was
connected to a new box containing two new Arduino Uno boards
(Figure~\ref{figura:Engine_room_Arduino_Box}).

\begin{figure}[H]
\centering
\includegraphics[width=0.8\textwidth]{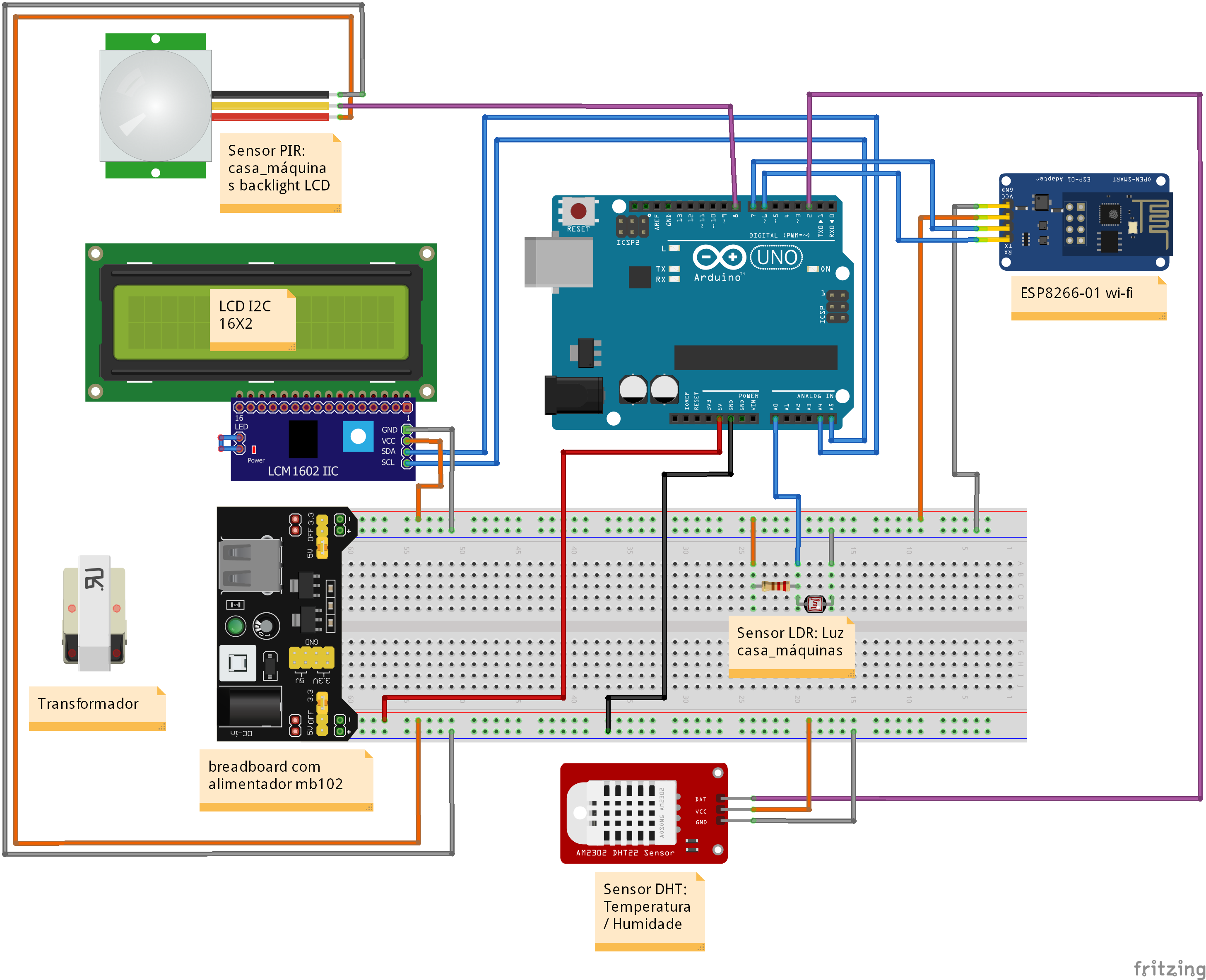}
\caption{\color{Gray} \textbf{Pool area -- hardware connection
    diagram.}}
\label{figura:Pool_area_Hardware_connection_diagram}
\end{figure}

\begin{figure}[h]
\centering
\includegraphics[width=.5\textwidth]{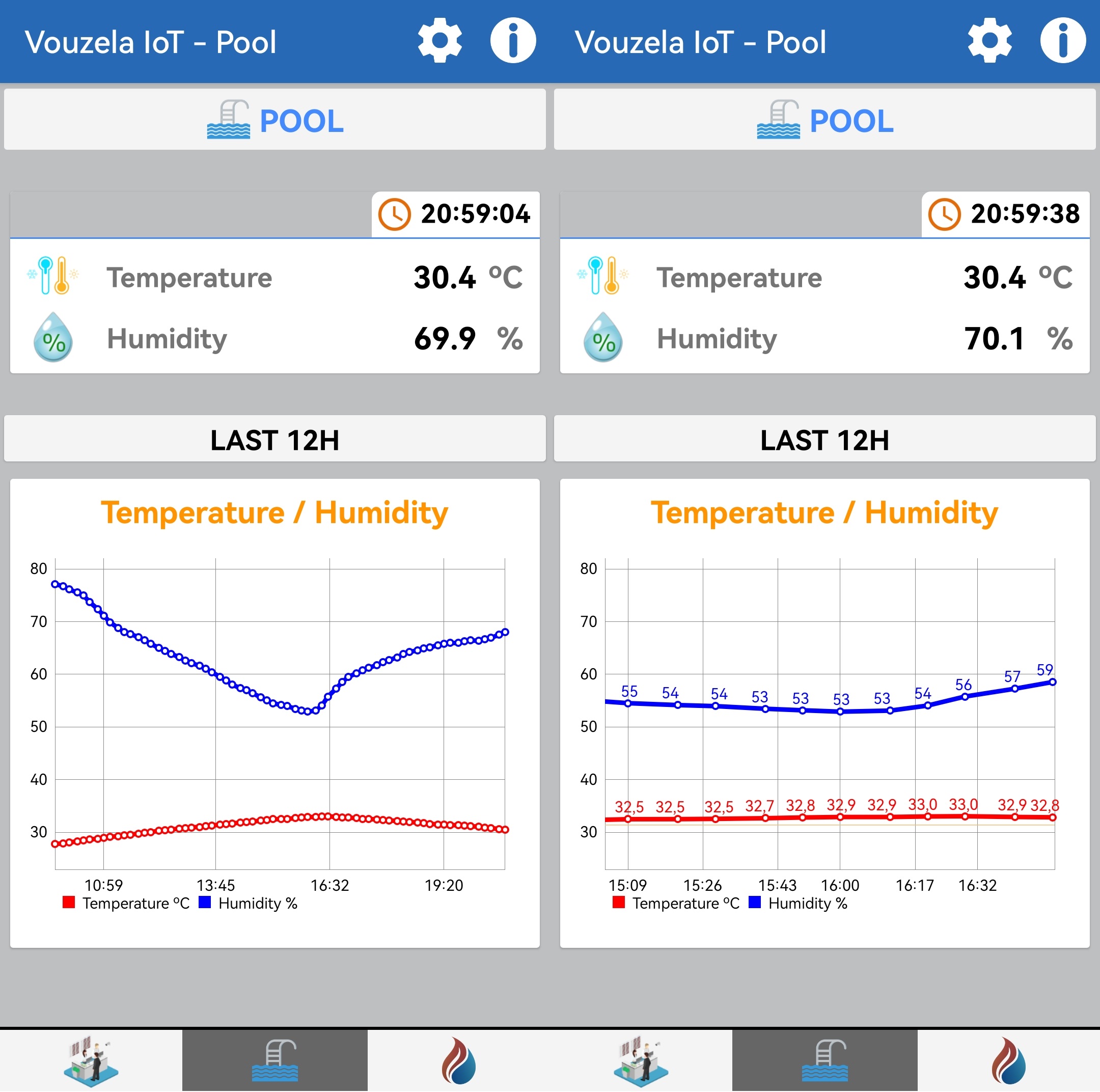}
\caption{\color{Gray} \textbf{Android -- temperature and humidity (Pool
    area).}}
\label{figura:Android_TempHum_Pool_area}
\end{figure}

Just like the Arduino board in the Reception area, the Arduino Uno
boards now installed in a new box in the engine room
(Figure~\ref{figura:Engine_room_Arduino_Box}) also have an independent
power supply, in order to guarantee the appropriate voltage for each
microcontroller, sensor, and actuator installed, both for the pool area
and for the engine room, which includes the boiler.

Next to the boxes where the Arduinos are installed, there are always
diagrams of the connections made to each board, so that the service
manager can correct the connections if necessary. In particular, on the
board where the temperature and relative humidity sensor is connected,
which collects data from the pool area, a motion sensor is also
connected to detect movement in the engine room. If there are any
records outside of working hours, a notification will be sent via the
mobile application to the responsible public service employee. In
addition to the motion sensor, a light sensor is also connected to the
same Arduino board, in order to inform, through the Android application,
whether the engine room light is on outside working hours. Monitoring
this parameter will help reduce energy consumption.
Figure~\ref{figura:Pool_area_Hardware_connection_diagram} shows the
hardware connection diagram for the Arduino board responsible for
collecting data from the pool. In the Android application
(Figure~\ref{figura:Android_TempHum_Pool_area}), in addition to reading
the data every 5 seconds on the temperature and relative humidity in the
environment around the pool, each point in the graph represents the
average of the data recorded every 10 minutes, ensuring that the graph
is not overwhelmed by excessive data. This way, those responsible can
easily observe the trends in temperature and relative humidity variation
over 12 hours and obtain a clear visual perspective of the environment
in the pool area, in order to monitor changing conditions and take
appropriate measures when necessary.

% ---------------------------------------------------------------
\subsubsection{Engine Room}

\begin{figure}[H]
\centering
\includegraphics[width=0.8\textwidth]{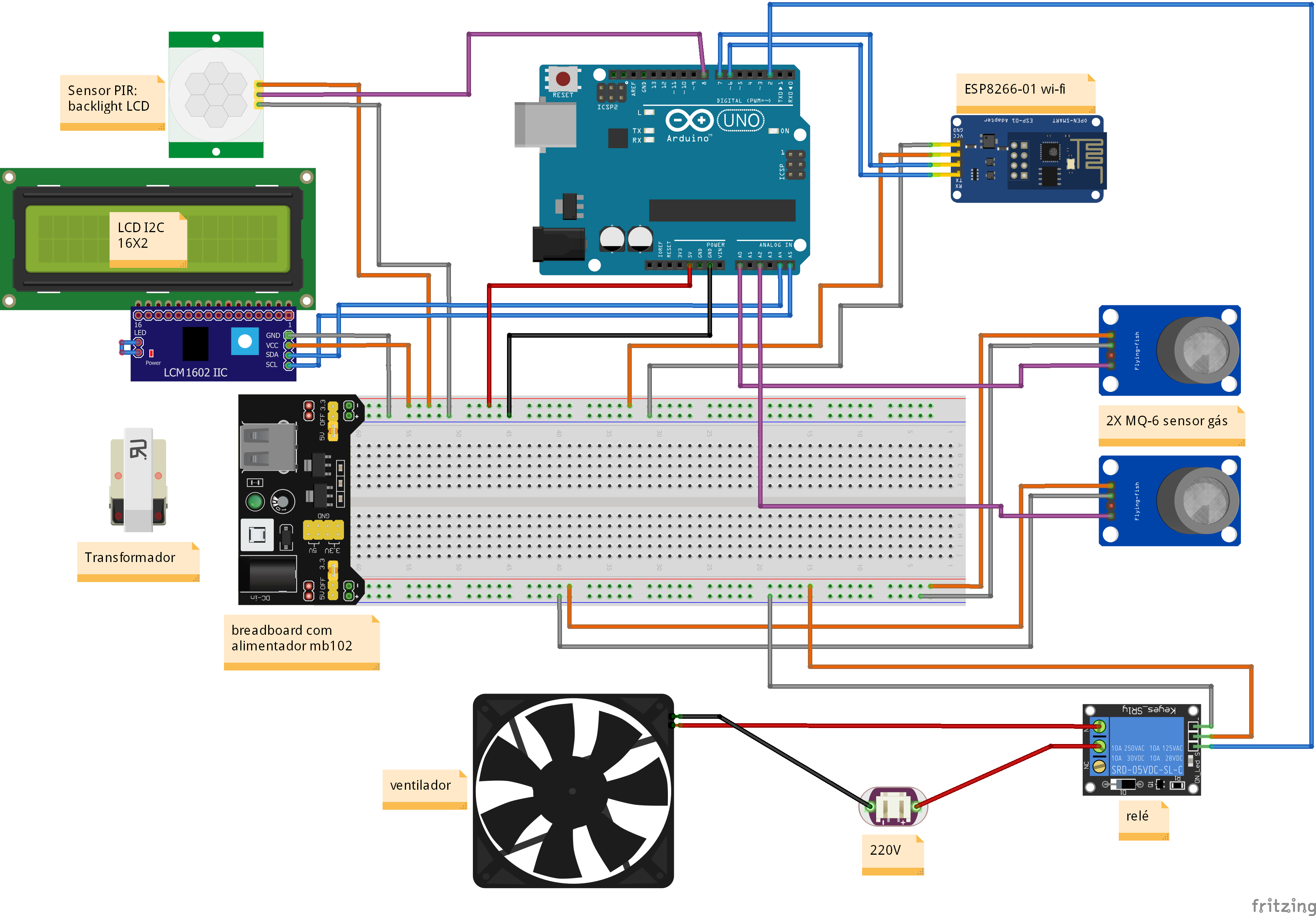}
\caption{\color{Gray} \textbf{Engine room -- hardware connection
    diagram.}}
\label{figura:Engine_room_Hardware_connection_diagram}
\end{figure}

\begin{figure}[ht]
  \centering
  \begin{subfigure}{.415\textwidth}
    \centering
    \includegraphics[width=\textwidth]{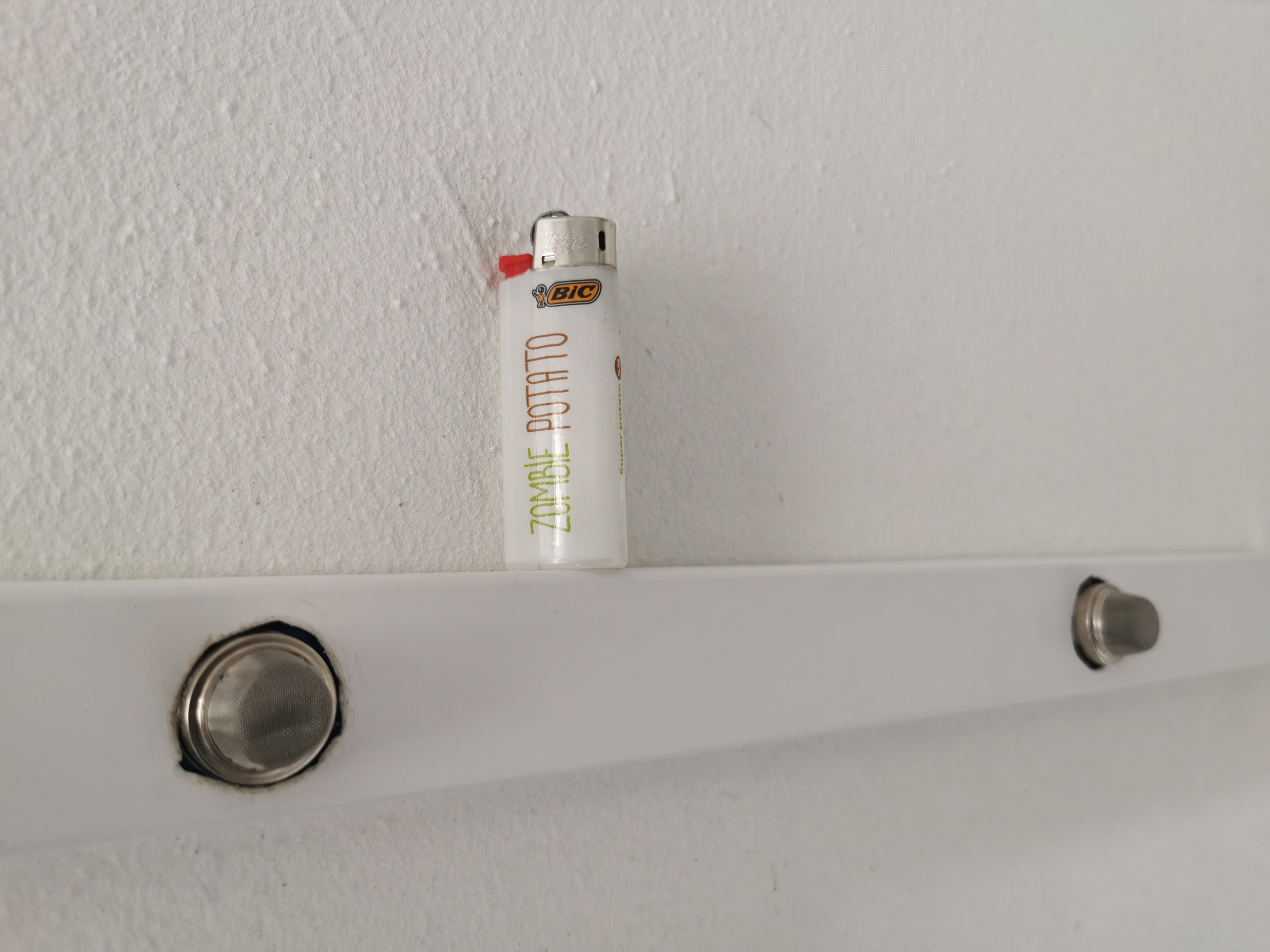}
    \caption{}
    \label{fig:Engine_room_Propane_gas_sensor}
  \end{subfigure}
  \qquad
  \begin{subfigure}{.4\textwidth}
    \centering
    \includegraphics[width=\textwidth]{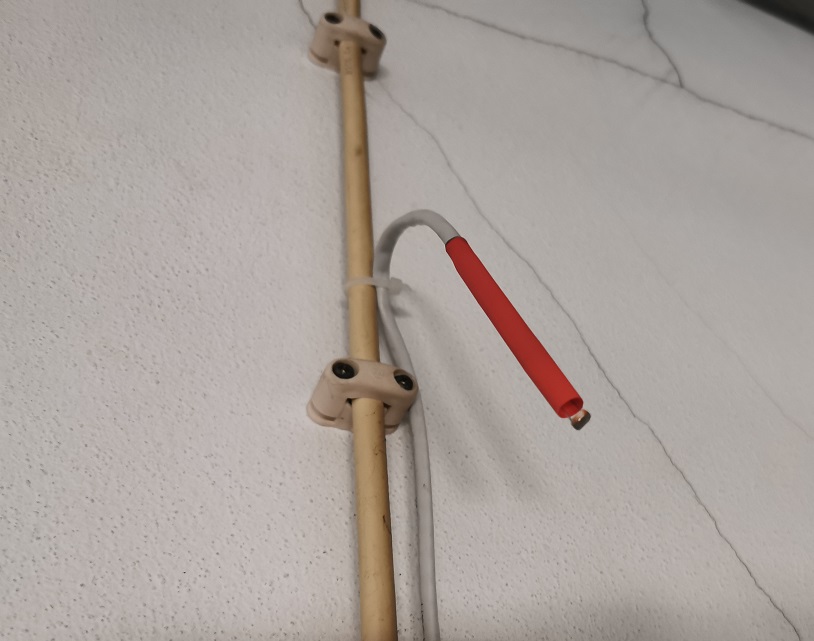}
    \caption{}
    \label{figura:LDR_pool_area}
  \end{subfigure}
  \caption{\color{Gray} \textbf{Sensors connected to the Arduino
    responsible for sending Engine Room values.}
    (a)~Gas sensor in the boiler area;
    (b)~LDR sensor.}
  \label{fig:gas_ldr_engroom}
\end{figure}

Still in the Engine Room, inside the same box, another Arduino Uno
board was installed to perform various monitoring and control functions.
This board is connected to propane gas detection sensors. Two MQ-6
sensors were connected to the Arduino Uno to detect any possible
propane gas leakage in the boiler space, as shown in
Figure~\ref{fig:Engine_room_Propane_gas_sensor} and according to the
hardware connection diagram proposed in
Figure~\ref{figura:Engine_room_Hardware_connection_diagram}.

These sensors use a metal oxide material which has electrical
conductivity properties that vary in response to the presence of
specific gases~\cite{ajiboye}.

When exposed to certain gases, metal oxides change their conductivity,
allowing the detection and quantification of the concentration of gases
present. Specifically, the MQ-6 sensors (Figure~\ref{fig:gassensor})
are highly sensitive to flammable gases, such as propane, and if any
abnormal concentration is detected, the Arduino Uno can trigger alarms
and take appropriate safety measures, such as stopping the gas supply or
turning on an extractor. Two gas detection sensors were programmed and
installed in order to mitigate possible false alarms; that is, if the
values of just one sensor are, for no apparent reason, greater than the
predefined value on the microcontroller, no alerts are triggered and the
relay actuator port is not activated. For this project, a relay was
connected, as an actuator, to the Arduino board, in order to allow
connection of an external gas extractor, a siren, or other device that
the person responsible for the service chooses to install. When the
relay status changes, a notification is also sent to the application as
well as to an email address defined for this purpose.

\begin{figure}[h]
  \centering
  \begin{subfigure}{.3\textwidth}
    \centering
    \includegraphics[width=\textwidth]{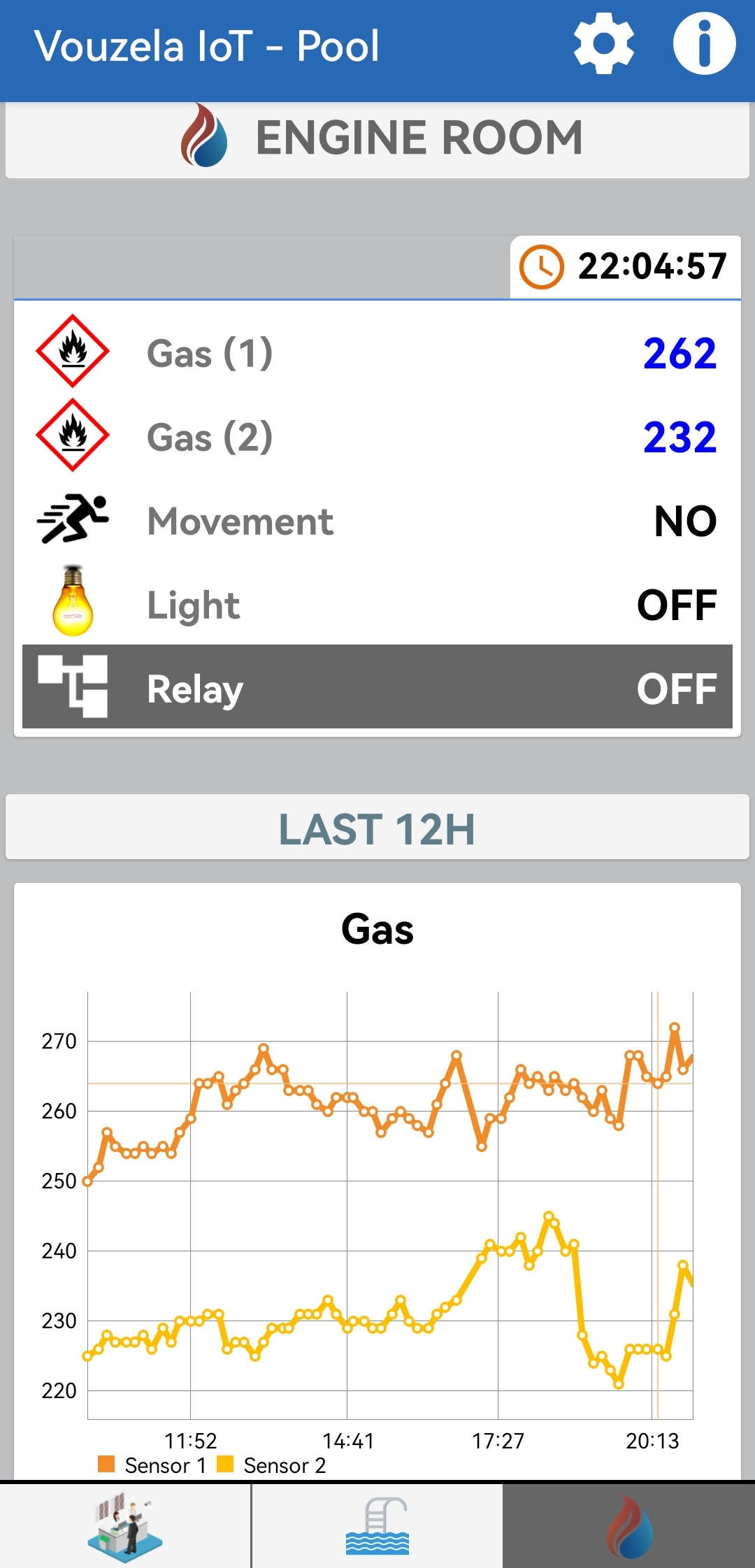}
    \caption{}
    \label{figura:Android_Gas_sensors_Engine_room}
  \end{subfigure}%
  \qquad
  \begin{subfigure}{.3\textwidth}
    \centering
    \includegraphics[width=\textwidth]{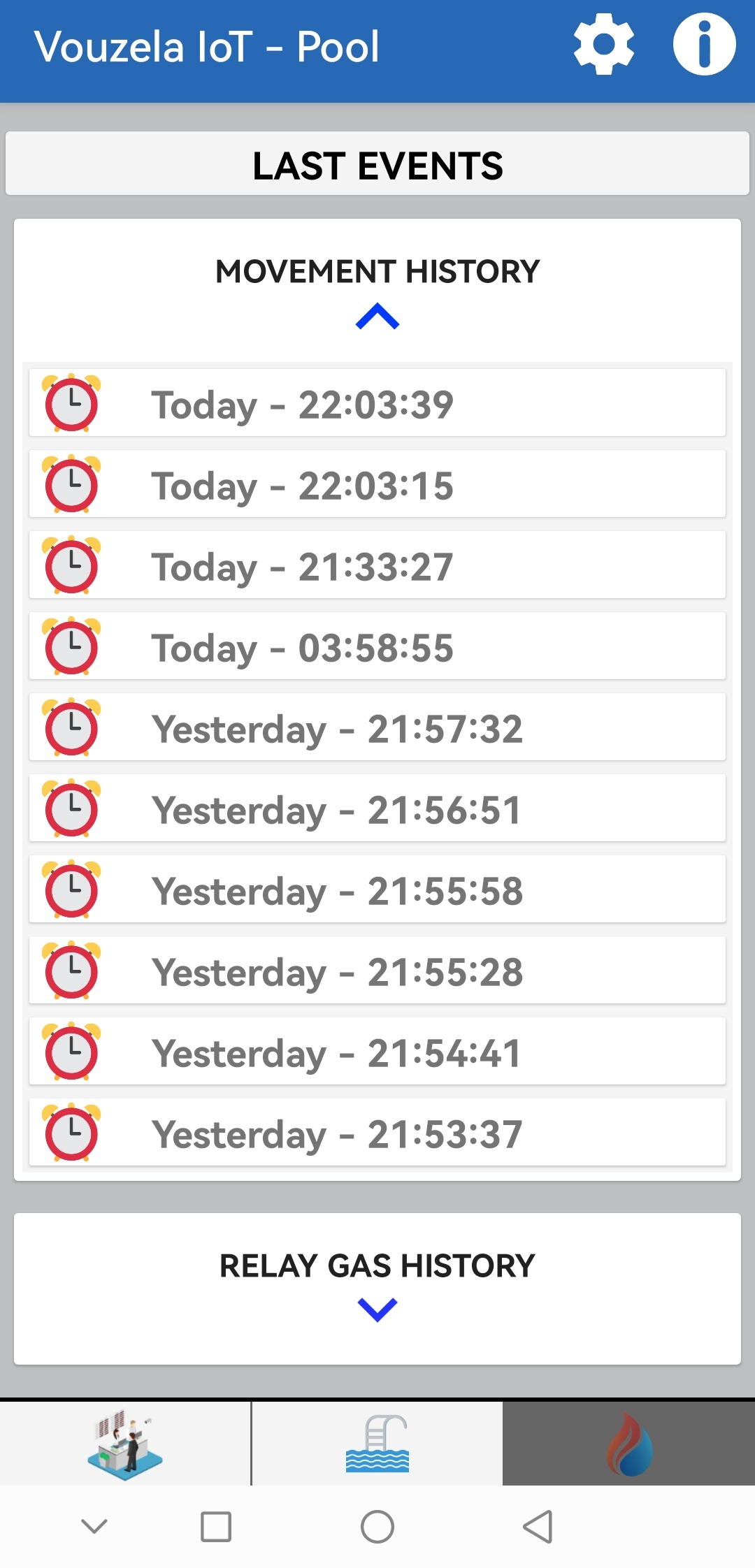}
    \caption{}
    \label{figura:Android_Movement_Engine_room}
  \end{subfigure}
  \caption{\color{Gray} \textbf{Android mobile application for the
    Engine Room.}
    (a)~Gas sensor parameters in real time and the gas variation graph
    over the last 12 hours;
    (b)~last movement events.}
  \label{fig:android_engine_room}
\end{figure}

With regard to the presentation of data in the Android application, the
interface relating to the Engine Room (where the boiler is included)
shows, in Figure~\ref{figura:Android_Gas_sensors_Engine_room}, the
sensor values over the last 12 hours and, equally important, the record
of the last 10 movements in that area, where there is an exit door to
the outside (Figure~\ref{figura:Android_Movement_Engine_room}). It is
important to note that the number of movements presented in the Android
application can be easily changed through the web API, without
compromising the functioning of the implemented system.

% ---------------------------------------------------------------
\section{Data Analysis and Discussion}\label{Data Analysis and Discussion}

This section presents the achieved results. Exploratory data analysis
will allow a better understanding of the home automation system and will
provide valuable information for decision-making and future
improvements. Temperature and relative humidity, as well as carbon
dioxide, will be analysed in more detail, as they are considered most
relevant by those responsible, especially in the pool and reception
areas. These parameters provide information to evaluate and monitor
environmental comfort, as well as the health of users and employees,
essential factors for the proper functioning of the public service.
After extracting the data from the InfluxDB database into files in CSV
(Comma-Separated Values) format to be processed using the Python
programming language, the time frame for analysis was defined as 5th to
25th June 2023, a period of 21 days of normal operation of the services
provided by the Vouzela municipal swimming pool.

% ---------------------------------------------------------------
\subsection{Carbon Dioxide (CO2)}

\begin{figure}[h]
\centering
\includegraphics[width=0.6\textwidth]{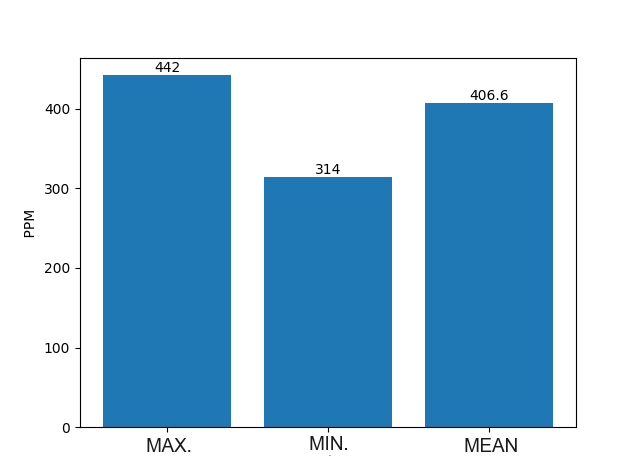}
\caption{\color{Gray} \textbf{Mean carbon dioxide over 21 days
    (Reception area).}}
\label{figura:Mean_CO2_21_days}
\end{figure}

Carbon dioxide is, naturally, a parameter to be analysed, as it is
directly related to the health of users. At certain times of the day
there is a greater number of people sharing the same air in the
reception area, and this led us to explore the data collected. Knowing
that, on a global scale, the average carbon dioxide concentration in
the atmosphere is 420~parts per million (ppm), according to the National
Oceanic and Atmospheric Administration (NOAA) of the United
States~\cite{noaa}, and the maximum acceptable level inside buildings is
650~ppm above the outdoor level~\cite{Afonso2018}, we can see in
Figure~\ref{figura:Mean_CO2_21_days} that the maximum value reached in
the reception area was 442~ppm, which does not justify an immediate
investment in air recirculation equipment. However, it is important to
highlight that the observed values represent only a period of 21 days in
the month of June, totalling 654,054 CO2 records.

% ---------------------------------------------------------------
\subsection{Temperature and Relative Humidity -- Reception Area}

\begin{figure}[H]
\centering
\includegraphics[width=0.6\textwidth]{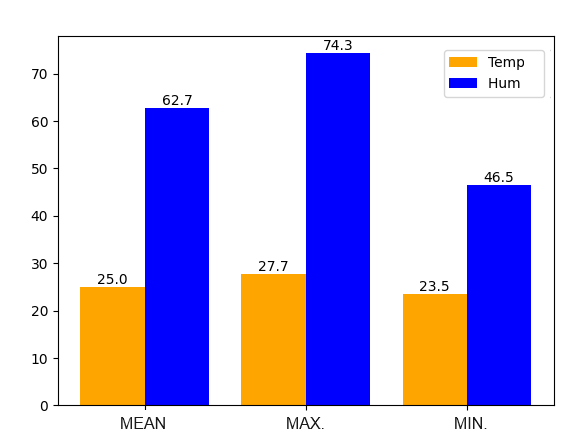}
\caption{\color{Gray} \textbf{Mean temperature and relative humidity
    over 21 days (Reception area).}}
\label{figura:Mean_TempHum_21_days_Reception}
\end{figure}

In the Reception area of the municipal swimming pool, an analysis of
data collected over a period of 21 days reveals important information
about temperature and relative humidity. In
Figure~\ref{figura:Mean_TempHum_21_days_Reception} we can see that the
average temperature recorded was 25\,$^{\circ}$C, reaching a maximum of
27.7\,$^{\circ}$C and a minimum of 23.5\,$^{\circ}$C. Relative humidity
had an average value of 62.7\%, reaching a maximum of 74.3\% and a
minimum of 46.5\%. This information is relevant to understanding the
environmental conditions in the reception space and, given the
temperature recorded, justifies the acquisition of air conditioning
equipment in order to offer greater comfort to users and employees, as
the temperature of workplaces should, as far as possible, range between
18\,$^{\circ}$C and 22\,$^{\circ}$C, except in certain weather
conditions in which it may reach 25\,$^{\circ}$C, and humidity between
50\% and 70\%, according to paragraph~a), point~1 of Article~11 of
Decree-Law 243/86, of 20 August~\cite{decretoLei}, which approves the
General Regulation on Occupational Hygiene and Safety in Commercial,
Office and Service Establishments in Portugal.

% ---------------------------------------------------------------
\subsection{Temperature and Relative Humidity -- Pool Area}

\begin{figure}[H]
\centering
\includegraphics[width=0.6\textwidth]{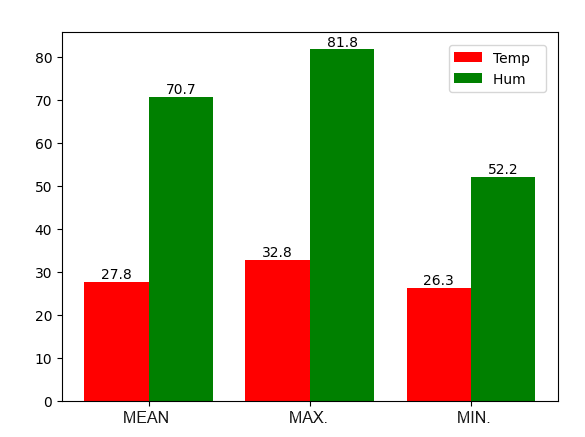}
\caption{\color{Gray} \textbf{Mean temperature and relative humidity
    over 21 days (Pool area).}}
\label{figura:Mean_TempHum_21_days_Pool}
\end{figure}

During the same 21-day period, a notable discrepancy in thermal
conditions was identified in the pool area. Analysis of the ambient
temperature surrounding the pool revealed a mean of
27.8\,$^{\circ}$C (Figure~\ref{figura:Mean_TempHum_21_days_Pool}),
contrasting with the pool water temperature, consistently maintained at
28.8\,$^{\circ}$C through daily manual and automatic measurements. This
deviation represents non-compliance with Portuguese Standard (NP)
4542:2017~\cite{norma4542}, which recommends a temperature difference
of 2\,$^{\circ}$C to 4\,$^{\circ}$C between ambient and water
temperatures. We emphasise that this issue requires attention not only
from a regulatory standpoint but also to ensure the well-being and
satisfaction of pool users. Considering the nature of this matter, it
is advisable to implement adjustments to align the conditions with the
established recommendations.

% ---------------------------------------------------------------
\section{Conclusion}\label{Conclusion}

The main objective of this project was the implementation of a home
automation system in the municipal swimming pool of Vouzela, aiming to
enhance the management, monitoring, and control of various environmental
and safety parameters in the space. Throughout the project development,
described in this paper, we addressed the key components of the system,
ranging from the hardware employed -- such as microcontrollers
integrated into Arduino boards, various sensors, and actuators -- to the
tools and technologies described in this document. Through the
integration of hardware components, communication was established
between devices, the API, and the Android mobile application, enabling
real-time data collection and processing. Data analysis was conducted
using the online Colab tool~\cite{colab}, allowing detailed exploration
through the Python programming language and associated libraries for
data processing and presentation.

From the data analysis, it was found that the maximum CO2 value
recorded in the Reception area was 442~ppm, which does not justify an
immediate investment in air recirculation equipment. However, it is
important to note that the observed values represent only a 21-day
period in June, during which over 1,700 door openings were recorded,
inevitably facilitating air recirculation. In addition, still in the
reception area, the average recorded temperature was 25\,$^{\circ}$C,
reaching a maximum of 27.7\,$^{\circ}$C and a minimum of
23.5\,$^{\circ}$C, justifying the acquisition of an air conditioning
system to provide greater comfort to users. Further analysis of
temperature and humidity data from the pool area revealed a discrepancy,
prompting alterations to the programming of the controllers monitoring
that space, as the temperature difference between the ambient air and
the pool water was less than 1\,$^{\circ}$C, whereas Portuguese
Standard (NP) 4542:2017, issued by the Portuguese Institute for Quality
(IPQ), indicates that the ambient temperature should be between
2\,$^{\circ}$C and 4\,$^{\circ}$C above the water temperature in the
pool.

During the system implementation, we faced technical and configuration
challenges, which were overcome with dedication, research, and
teamwork. The collaboration of the members of the Municipality of
Vouzela was crucial for the success of this project. Through this work,
the benefits of applying home automation in the management of public
spaces were demonstrated, providing a safer, more efficient, and more
comfortable environment for users. Additionally, the use of modern
technologies and the analysis of collected data allowed for a more
informed decision-making process and the implementation of preventive
measures to ensure the quality and sustainability of the facilities.

% ---------------------------------------------------------------
\section*{Acknowledgements}

The development team gratefully acknowledges the Municipality of Vouzela
for its valuable collaboration in granting access to the Municipal
Swimming Pool facilities, which were instrumental in the development and
validation of this project.

%\nolinenumbers

\bibliography{library}

@INPROCEEDINGS{shah,
  author={Shah, Sajjad Hussain and Yaqoob, Ilyas},
  booktitle={2016 IEEE Smart Energy Grid Engineering (SEGE)}, 
  title={A survey: Internet of Things (IOT) technologies, applications and challenges}, 
  year={2016},
  volume={},
  number={},
  pages={381-385},
  doi={10.1109/SEGE.2016.7589556}}

@article{alves2022domotica,
  title={Dom{\'o}tica: Estudo da contribu{\c{c}}{\~a}o da Aautoma{\c{c}}{\~a}o residencial para a acessibilidade de portadores de defeci{\^e}ncia f{\'i}sica},
  author={Alves, Rafael Calderan and Florian, Fabiana and Farina, Renata Mirella},
  journal={Recima21-Revista Cient{\'\i}fica Multidisciplinar-ISSN 2675-6218},
  volume={3},
  number={12},
  pages={e3122299--e3122299},
  year={2022}
}

@book{bolzani2004residencias,
  title={Resid{\^e}ncias Inteligentes},
  author={Bolzani, C.A.M.},
  isbn={9788588325258},
  url={https://books.google.pt/books?id=tgTlPE10u68C},
  year={2004},
  publisher={Ed. Livraria da F{\'\i}sica}
}

@InProceedings{gupta,
author="Gupta, Malvika
and Singh, Shweta",
editor="Gao, Xiao-Zhi
and Kumar, Rajesh
and Srivastava, Sumit
and Soni, Bhanu Pratap",
title="A Survey on the ZigBee Protocol, It's Security in Internet of Things (IoT) and Comparison of ZigBee with Bluetooth and Wi-Fi",
booktitle="Applications of Artificial Intelligence in Engineering",
year="2021",
publisher="Springer Singapore",
address="Singapore",
pages="473--482",
isbn="978-981-33-4604-8"
}

@article{misra2018iot,
  title={An IoT-based waste management system monitored by cloud},
  author={Misra, Debajyoti and Das, Gautam and Chakrabortty, Triankur and Das, Debaprasad},
  journal={Journal of Material Cycles and Waste Management},
  volume={20},
  number={3},
  pages={1574--1582},
  year={2018},
  publisher={Springer}
}

@inproceedings{ramphela2020internet,
  title={Internet of things (IoT) integrated data center infrastructure monitoring system},
  author={Ramphela, Mahlatsi Kgabo Jackson and Owolawi, Pius Adewale and Mapayi, Temitope and Aiyetoro, Gbolahan},
  booktitle={2020 International Conference on Artificial Intelligence, Big Data, Computing and Data Communication Systems (icABCD)},
  pages={1--6},
  year={2020},
  organization={IEEE}
}

@inproceedings{saha2017data,
  title={Data centre temperature monitoring with ESP8266 based Wireless Sensor Network and cloud based dashboard with real time alert system},
  author={Saha, Saraswati and Majumdar, Anupam},
  booktitle={2017 Devices for Integrated Circuit (DevIC)},
  pages={307--310},
  year={2017},
  organization={IEEE}
}

@INPROCEEDINGS{bing,
  author={Bing, Kang and Fu, Liu and Zhuo, Yun and Yanlei, Liang},
  booktitle={2011 2nd International Conference on Intelligent Control and Information Processing}, 
  title={Design of an Internet of Things-based smart home system}, 
  year={2011},
  volume={2},
  number={},
  pages={921-924},
  doi={10.1109/ICICIP.2011.6008384}}

@electronic{phpmailer,
  title         = "PHP Mailer",
  url           = "https://github.com/PHPMailer/PHPMailer",
  year          = "2022",
  note          = "visited on 2022-04-19",
}

@misc{sparkpostr,
  title         = "sparkpost",
  url           = "https://www.sparkpost.com/",
  year          = "2022",
  note          = "visited on 2022-04-20",
}

@misc{esp01,
  title         = "MasterWalker Blog: Upgrade de Firmware do WiFi ESP8266 ESP-01 através do Arduino",
  url           = "https://blogmasterwalkershop.com.br/embarcados/esp8266/upgrade-de-firmware-do-wifi-esp8266-esp-01-atraves-do-arduino-e-conversor-usb-serial",
  year          = "2022",
  note          = "visited on 2023-03-20",
}

@misc{esp01toapi,
  title         = "MasterWalker Blog: Como usar com Arduino – Módulo WiFi ESP8266 ESP-01",
  url           = "https://blogmasterwalkershop.com.br/arduino/como-usar-com-arduino-modulo-wifi-esp8266-esp-01",
  year          = "2022",
  note          = "visited on 2023-03-22",
}

@misc{dht22,
  title         = "Gravity: DHT22 Temperature \& Humidity Sensor",
  url           = "https://www.dfrobot.com/product-1102.html",
  note          = "visited on 2023-04-10",
}

@misc{pir,
  title         = "How PIRs Work",
  url           = "https://learn.adafruit.com/pir-passive-infrared-proximity-motion-sensor/how-pirs-work",
  year          = "2023",
  note          = "visited on 2023-03-20",
}

@misc{pirspecs,
  title         = "PIR Sensors Guide with Arduino Programming for motion detection",
  url           = "https://www.electronicwings.com/sensors-modules/pir-sensor",
  year          = "2023",
  note          = "visited on 2023-04-06",
}

@misc{ldr,
  title         = "MasterWalker Blog: Como usar com Arduino – Fotoresistor (Sensor) LDR 5MM",
  url           = "https://blogmasterwalkershop.com.br/arduino/como-usar-com-arduino-fotoresistor-sensor-ldr-5mm",
  note          = "visited on 2023-04-12",
}

@misc{mc38,
  title         = "Arduino Get Started: Door sensor",
  url           = "https://arduinogetstarted.com/tutorials/arduino-door-sensor",
  year          = "2023",
  note          = "visited on 2023-04-13",
}

@misc{rele,
  title         = "MasterWalker Blog: Como usar com Arduino – Módulo Relé 5V 1 Canal",
  url           = "https://blogmasterwalkershop.com.br/arduino/como-usar-com-arduino-modulo-rele-5v-1-canal",
  note          = "visited on 2023-04-15",
}

@misc{lcd,
  title         = "Arduino Get Started: LCD I2C",
  url           = "https://arduinogetstarted.com/tutorials/arduino-lcd-i2c",
  year          = "2023",
  note          = "visited on 2023-04-15",
}

@misc{influxdb,
  title         = "InfluxDB - It's About Time.",
  url           = "https://www.influxdata.com/",
  year          = "2023",
  note          = "visited on 2023-04-18",
}

@misc{influx1,
  title         = "InfluxDB v1 documentation",
  url           = "https://docs.influxdata.com/influxdb/v1/",
  year          = "2023",
  note          = "visited on 2023-04-15",
}

@misc{influxphp,
  title         = "InfluxDB PHP Library",
  url           = "https://github.com/influxdata/influxdb-php",
  year          = "2023",
  note          = "visited on 2023-04-15",
}

@misc{onesignaldoc,
  title         = "Onesignal: Android SDK Setup",
  url           = "https://documentation.onesignal.com/docs/android-sdk-setup",
  year          = "2023",
  note          = "visited on 2023-05-02",
}

@misc{onesignalgithub,
  title         = "Github: OneSignal Android Push Notification Plugin",
  url           = "https://github.com/OneSignal/OneSignal-Android-SDK",
  year          = "2023",
  note          = "visited on 2023-05-02",
}

@misc{fail2bangithub,
  title         = "Github: Fail2Ban - ban hosts that cause multiple authentication errors",
  url           = "https://github.com/fail2ban/fail2ban",
  year          = "2022",
  note          = "visited on 2023-05-10",
}

@article{naqvi2017time,
  title={Time series databases and influxdb},
  author={Naqvi, Syeda Noor Zehra and Yfantidou, Sofia and Zim{\'a}nyi, Esteban},
  journal={Studienarbeit, Universit{\'e} Libre de Bruxelles},
  volume={12},
  year={2017}
}

@article{liu2016green,
  title={Green data center with IoT sensing and cloud-assisted smart temperature control system},
  author={Liu, Qiang and Ma, Yujun and Alhussein, Musaed and Zhang, Yin and Peng, Limei},
  journal={Computer Networks},
  volume={101},
  pages={104--112},
  year={2016},
  publisher={Elsevier}
}

@Article{valente,
AUTHOR = {Valente, Antonio and Costa, Carlos and Pereira, Leonor and Soares, Bruno and Lima, José and Soares, Salviano},
TITLE = {A LoRaWAN IoT System for Smart Agriculture for Vine Water Status Determination},
JOURNAL = {Agriculture},
VOLUME = {12},
YEAR = {2022},
NUMBER = {10},
ARTICLE-NUMBER = {1695},
URL = {https://www.mdpi.com/2077-0472/12/10/1695},
ISSN = {2077-0472},
DOI = {10.3390/agriculture12101695}
}

@inproceedings{flores-cortez,
author = {Flores-Cortez, Omar and Cortez, Ronny and Rosa, Veronica},
year = {2022},
month = {07},
pages = {},
title = {Implementación de un sistema IoT de bajo costo para el monitoreo de la calidad del aire en El Salvador}
}

@mastersthesis{SanzGresa2021,
  author = {Sanz Gresa, B.},
  title = {Diseño e implementación de un sistema de mantenimiento de una piscina basado en Arduino},
  school = {Universitat Politècnica de València},
  year = {2021},
  type = {Grado},
  url = {http://hdl.handle.net/10251/174697}
}

@InProceedings{rocha2023,
author="Rocha, J{\'u}lio
and Lucas, Marco
and Figueiredo, Ricardo
and Henriques, Jo{\~a}o
and Bernardo, Marco V.
and Wanzeller, Cristina
and Caldeira, Filipe",
editor="de la Iglesia, Daniel H.
and de Paz Santana, Juan F.
and L{\'o}pez Rivero, Alfonso J.",
title="A Cost-Effective Framework for Monitoring Disaster Recovery Infrastructures",
booktitle="New Trends in Disruptive Technologies, Tech Ethics and Artificial Intelligence",
year="2023",
publisher="Springer International Publishing",
address="Cham",
pages="201--211",
isbn="978-3-031-14859-0"
}

@article{satapathy2018arduino,
  title={Arduino based home automation using Internet of things (IoT)},
  author={Satapathy, Lalit Mohan and Bastia, Samir Kumar and Mohanty, Nihar},
  journal={International Journal of Pure and Applied Mathematics},
  volume={118},
  number={17},
  pages={769--778},
  year={2018}
}

@article{Saptiani_2019,
doi = {10.1088/1742-6596/1280/2/022058},
url = {https://dx.doi.org/10.1088/1742-6596/1280/2/022058},
year = {2019},
month = {nov},
publisher = {IOP Publishing},
volume = {1280},
number = {2},
pages = {022058},
author = {P Saptiani and M H Aziz and M Iriyanti and A Aminudin},
title = {The electrical properties characterization of MG-811 gas sensor toward the temperature alteration of soil testing chamber},
journal = {Journal of Physics: Conference Series},
}

@article{ajiboye,
author = {Ajiboye, Aye and F., Opadiji and Yusuf, Abdulrahman and O., Popoola},
year = {2021},
month = {04},
pages = {575},
title = {Analytical determination of load resistance value for MQ-series gas sensors: MQ-6 as case study},
volume = {19},
journal = {TELKOMNIKA (Telecommunication Computing Electronics and Control)},
doi = {10.12928/telkomnika.v19i2.17427}
}

@misc{noaa,
  title         = "Global Monitoring Laboratory",
  url           = "https://gml.noaa.gov/ccgg/trends/global.html",
  year          = "2023",
  note          = "visited on 2023-09-04",
}

@misc{uno,
  title         = "Arduino UNO",
  url           = "https://www.javatpoint.com/arduino-uno",
  year          = "2021",
  note          = "visited on 2023-03-16"
}

@misc{mega,
  title         = "Arduino Mega",
  url           = "https://www.javatpoint.com/arduino-mega",
  year          = "2021",
  note          = "visited on 2023-03-16"
}

@misc{gson,
  title         = "Gson",
  url           = "https://github.com/google/gson",
  year          = "2022",
  note          = "visited on 2023-03-14",
}

@misc{okHttp,
  title         = "OkHttp",
  url           = "https://square.github.io/okhttp/",
  year          = "2022",
  note          = "visited on 2023-03-14"
}

@misc{wifiesp,
  title         = "Arduino Libraries - WiFiEsp",
  url           = "https://www.arduino.cc/reference/en/libraries/wifiesp/",
  year          = "2017",
  note          = "visited on 2023-02-27"
}

@misc{colab,
  title         = "Google Colaboratory",
  url           = "https://colab.google/",
  note          = "visited on 2023-08-27"
}

@misc{lora,
  title         = "What is LoRaWAN Specification",
  url           = "https://lora-alliance.org",
  year          = "2022",
  note          = "visited on 2023-02-17"
}

@Article{Afonso2018,
AUTHOR = {Afonso, Clito and Gonçalves, Rui},
TITLE = {Qualidade do Ar por Renovação do Ar por Controlo Contínuo de CO2 Versus Injeção Contínua de Ar Novo: Consumos Energéticos Associados e Benefícios},
JOURNAL = {Revista da Associação Portuguesa de Análise Experimental de Tensões},
VOLUME = {30},
YEAR = {2018},
PAGES = {43-54},
URL = {https://hdl.handle.net/10216/118579},
ISSN = {1646-7078}
}

@online{decretoLei,
  title = {Decreto-Lei n.º 243/86, de 20 de agosto},
  author = {Assembleia da República},
  year = {1986},
  url = {https://diariodarepublica.pt/dr/detalhe/decreto-lei/243-1986-219080}
}

@online{norma4542,
  title = {Norma Portuguesa 4542:2017 - Piscinas: Requisitos de qualidade e tratamento da água para uso nos tanques},
  author = {Instituto Português da Qualidade},
  year = {2017},
  url = {https://www.apppiscinas.pt/uploads/JGoJULcnqym1YLhqhxcA.pdf}
}
\bibliographystyle{unsrt}

\end{document}